\documentclass[12pt,preprint]{aastex}
\usepackage{amsbsy}
\usepackage{amsmath}

\usepackage{epstopdf}
\usepackage{graphicx}
\shorttitle{Submillimeter Polarization of Galactic Clouds}
\shortauthors{Vaillancourt \& Matthews}

\begin{document}

\title{Submillimeter Polarization of Galactic Clouds:\\A Comparison of
  $350\,\micron$ and $850\,\micron$ Data}
\author{John E. Vaillancourt}
\affil{SOFIA Science Center, Universities Space Research Association,
  NASA Ames Research Center, Moffett Field, CA 94035;
  \email{jvaillancourt@sofia.usra.edu}}
\and
\author{Brenda C. Matthews}
\affil{Herzberg Institute, National Research Council of Canada, 5071
  W. Saanich Road, Victoria, BC V9E 2E7, Canada;
  \email{brenda.matthews@nrc-cnrc.gc.ca}}

\begin{abstract}
  The Hertz and SCUBA polarimeters, working at $350\,\micron$ and
  $850\,\micron$ respectively, have measured the polarized emission in
  scores of Galactic clouds.  Of the clouds in each dataset, 17 were
  mapped by both instruments with good polarization signal-to-noise
  ratios.  We present maps of each of these 17 clouds comparing the
  dual-wavelength polarization amplitudes and position angles at the
  same spatial locations.  In total number of clouds compared, this is
  a four-fold increase over previous work.  Across the entire data-set
  real position angle differences are seen between wavelengths. While
  the distribution of $\phi(850)-\phi(350)$ is centered near zero
  (near-equal angles), 64\% of data points with high polarization
  signal-to-noise ($P\ge3\sigma_p$) have
  $\vert\phi(850)-\phi(350)\vert>10\arcdeg$. Of those data with small
  changes in position angle ($\le 10\arcdeg$) the median ratio of the
  polarization amplitudes is $P(850)/P(350) = 1.7\pm0.6$.  This value
  is consistent with previous work performed on smaller samples and
  models which require mixtures of different grain properties and
  polarization efficiencies. Along with the polarization data we have
  also compiled the intensity data at both wavelengths; we find a
  trend of decreasing polarization with increasing 850-to-350
  $\micron$ intensity ratio.  All the polarization and intensity data
  presented here (1699 points in total) are available in electronic
  format.
\end{abstract}


\section{Introduction}


Observations of polarized radiation in the interstellar medium at
optical through millimeter wavelengths have been attributed to
extinction by, and emission from, interstellar dust grains (e.g.,
\citealt{hiltner1949b,hiltner1951,rhh88}).  In order to generate a net polarization the
grains must be aspherical and exhibit a relative net alignment of
their axes with one another and with the interstellar magnetic-field,
typically the shortest grain axis is parallel to the field (e.g.,
\citealt{dg51,alexreview1,alexreview2}).  At near-optical wavelengths
this polarizing dust screen causes the dichroic extinction of
background starlight with respect to the grain axes' different
cross-sections.  At far-infrared and longer wavelengths dominated by
grain emission rather than extinction the polarization results from
the axes' different emission cross-sections.
Due to the necessary role of magnetic fields in aligning the grains,
polarization observations have been used primarily to study
interstellar magnetic fields (e.g.,
\citealt{rhh88,fosalba2002,crutcher2004mim,archivescuba,pereyra2007}).
Specifically, the magnetic field morphology (projected onto the
plane-of-the-sky) is inferred from the polarization position
angles. (The field is typically parallel to the polarization angle in
the case of extinction and perpendicular in the case of emission.)
However, the physical properties of the grains themselves and their interaction
with the field are no less important than the field itself; many of
these properties can be inferred from the polarization amplitude.
For example, spectropolarimetry of background-starlight at
near-optical wavelengths has been used to measure the shapes of dust
grains \citep{shape}, make tests of grain alignment mechanisms (e.g.,
\citealt{whittet08,andersson07}), place limits on the size of aligned
grains (e.g., \citealt{km1995}), and measure the composition of the
aligned grains via polarized spectral lines (e.g.,
\citealt{whittet04}).


Polarimetry at submillimeter wavelengths was initially driven by the
desire to study the magnetic-field morphology of interstellar clouds.
As such, the amplitude of the polarization (typically $\sim$1--10\%),
and any wavelength dependence, was mostly secondary to
measurements of the polarization position angle.
Most studies of the polarization spectrum in the far-infrared and
submillimeter have thus relied on observations where the choice of
wavelength was made according to the availability of atmospheric
observing windows, not with the specific goal of studying any spectral
variation itself.  This mode of operation has resulted in a number of
databases at a few specific wavelengths including 60 and 100~$\micron$
\citep{archive}, $350\,\micron$ \citep{hertzarchive}, and
$850\,\micron$ \citep{scubaarchive}. Using a subset of the available
data, \citet{pspec} showed that the polarization spectrum across these
four wavelengths had a minimum near $350\,\micron$ (see also
\citealt{tenerife}).

The spectral structure observed in near-optical continuum polarimetry
(i.e., the ``Serkowski law''; \citealt*{serkowski75}) is the result of
a combination of properties: a) interstellar dust grains have typical
radii $a\sim0.1$--1~$\micron$ (e.g., \citealt*{mrn77}); b) the larger
grains are more efficiently aligned than the smaller grains; and c)
the grain sizes probed are of the same order as the observing
wavelengths $\lambda\sim a$ (e.g., \citealt{km1995}).  On the other
hand, at wavelengths in and beyond the far-infrared ($\lambda \gtrsim
50\,\micron$) all the above properties are the same save for the fact
that the grains are comparatively small ($\lambda \gg a$).  In this
case one expects no variation in the polarization spectrum.
Therefore, any model explaining the spectral structure observed by
\citet{pspec} requires multiple dust grain populations in which there
is a correlation between the efficiency with which the grains are
aligned and other properties related to the emitted radiation (i.e.,
grain size, temperature, emissivity).  We describe some simple models
in Section~\ref{sec-discussion}.

The initial studies of \citet{pspec} did not have sufficient data to
test such models.
\citet{mythesis} and \citet{omc1sharp} extended these datasets slightly, 
performing cloud-by-cloud comparisons as well as point-by-point
spectral comparisons within clouds.  While \citeauthor{pspec}'s
original result held-up under this more detailed analysis, the later
work was still limited to a small number of molecular clouds. The
recent compilation of large re-reduced datasets at $350\,\micron$ and
$850\,\micron$ presents the opportunity to further extend the sample
to point-by-point comparisons in additional Galactic molecular clouds.
In this work we present a comparison of submillimeter polarization data
at these wavelengths in a total of 17 clouds.

We present polarization maps of these 17 objects comparing the
polarization amplitude and position angle at the two wavelengths.
Section~\ref{sec-compare} highlights differences between the two
wavelengths in both angle and amplitude.  Changes in the angle may
help disentangle the magnetic field morphology along the line of sight
or extend maps to regions not observable at some wavelengths (e.g.,
\citealt{w3, kandori2007,li2009,quebec2009}), but we do not elaborate
on the position angle data presented here.  In
Section~\ref{sec-compare} we also compare the 850-to-350~$\micron$
polarization ratio on a point-by-point basis.  In
Section~\ref{sec-discussion} we compare both the absolute polarization
magnitudes, as well as the 850-to-350~$\micron$ polarization ratio,
with the 850-to-350~$\micron$ intensity ratio, compare it to previous
work, and briefly discuss grain alignment models consistent with the
data. All the polarization and intensity data presented in this paper
are available in machine readable tables in the electronic version of the
journal.

\section{Data Processing}

\begin{figure}
    \plotone{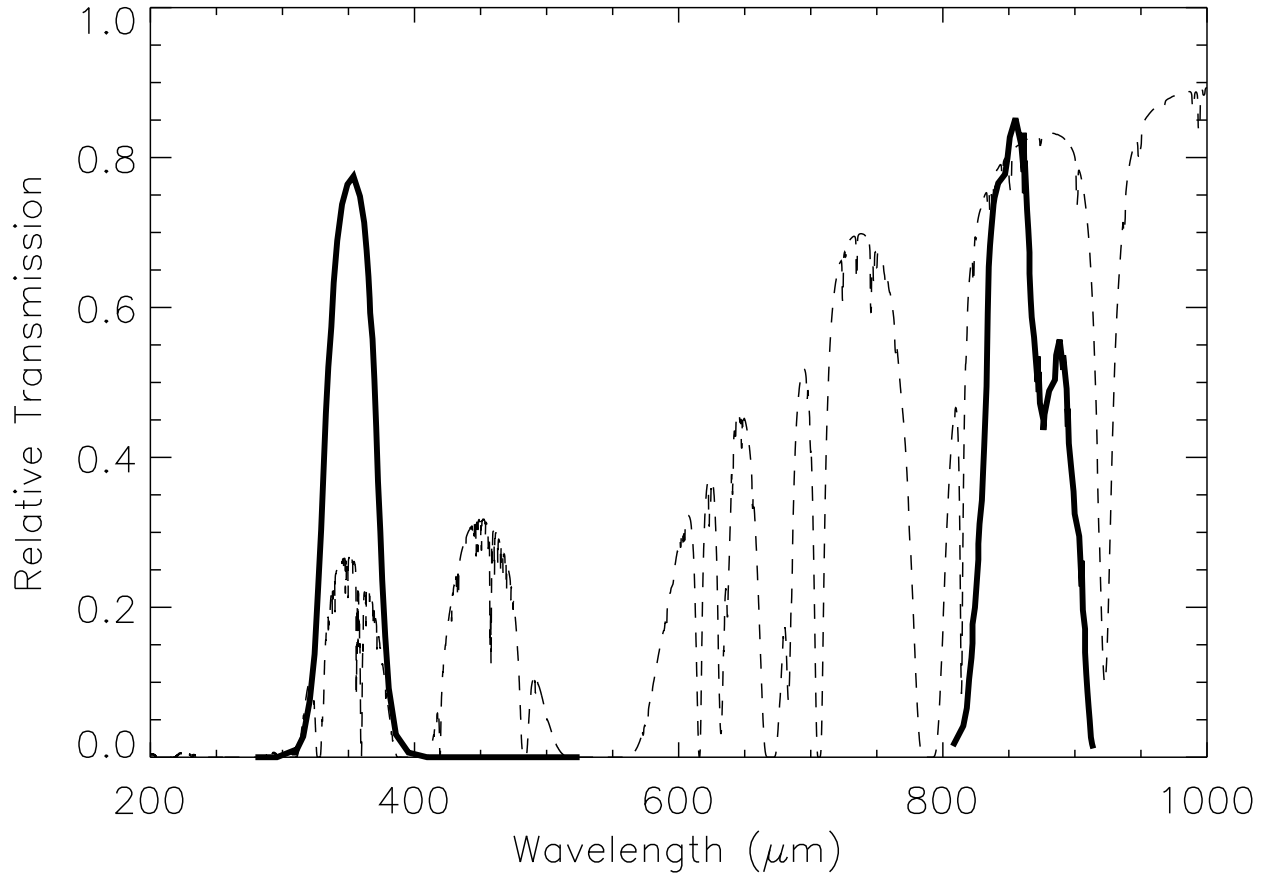}
    \caption{The Hertz $350\,\micron$ and SCUBA $850\,\micron$
      passbands (solid curves) are shown along with the typical
      atmospheric transmission on Mauna Kea (dashed line).  The
      atmospheric transmission is calculated for 1\,mm of precipitable
      water vapor using the CSO's web-based interface
      (http://www.submm.caltech.edu/cso/weather/atplot.shtml).  }
    \label{fig-filters}
\end{figure}

From $\sim$ 1995 to 2005 independent campaigns to map the polarization at
$350\,\micron$ and $850\,\micron$ were carried out by the Hertz
polarimeter at the Caltech Submillimeter Observatory (CSO) and the
SCUBA polarimeter at the James Clerk Maxwell Telescope (JCMT), both on
Mauna Kea, Hawaii.  The Hertz passband was chosen to match the
$350\,\micron$ atmospheric window while SCUBA's polarimeter operated
primarily at $850\,\micron$ (Figure \ref{fig-filters}).

\subsection{Spatial Resolution and Map Sampling}

\begin{figure}
    \plotone{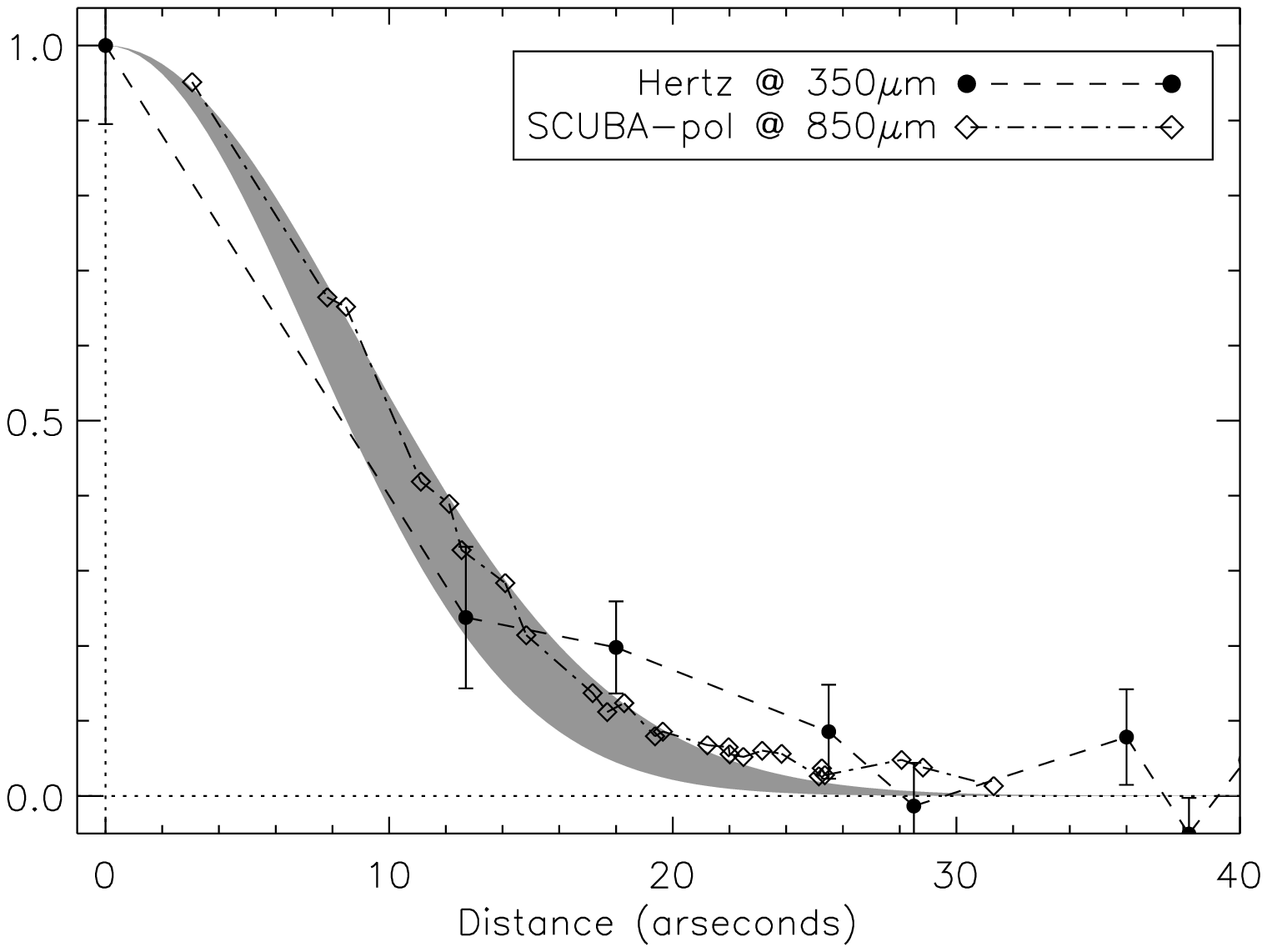}
    \caption{The Hertz $350\,\micron$ (circles and dashed line) and
      SCUBA-pol $850\,\micron$ (diamonds and dash-dotted line) beams
      were measured on Uranus in September 1997 and July 1998,
      respectively.  In both cases the disk's angular diameter was
      3\farcs7. These data represent the effective beam-widths after
      all data processing steps.  Gaussian beams with full-widths at
      half-maximum ranging from $17\arcsec$ -- $21\arcsec$ are shown
      for comparison (gray).  Error-bars on the Hertz data indicate
      the variation of the signal within an annulus centered at each
      radius \citep{hertz2}. Error-bars on the SCUBA-pol points were
      calculated from intensity uncertainties at each data pixel and are
      smaller than the plotted data points.}
    \label{fig-beams}
\end{figure}

The Hertz instrument, its observing strategy, and data analysis are
described in detail elsewhere \citep{kirby05,hertz2,hertz1,stokes}.
Here we briefly review the aspects relevant to the present work.
Hertz incorporates two separate detector arrays which simultaneously
measure the two orthogonal modes of linear polarization, modulated by
a cryogenic half-wave plate (HWP)\@. The $6\times6$ pixel$^2$ arrays
have pixel center-to-center spacings of $17\farcs8$ and a beamsize of
approximately $20\arcsec$ full-width at half-maximum (FWHM)\@.  The
observing strategy involves rotating the instrument to follow the
sky-rotation throughout the night, as well as steps of order the
array-size to map areas larger than the $2\arcmin\times 2\arcmin$
field-of-view. The rotation allowed a single pixel to continue
observing the same patch of sky throughout the night. Additionally,
step-sizes were typically chosen to be an integer-number of pixels;
only rarely were maps made with samples spaced more closely than the
$17\farcs8$ pixel pitch.
As a result, these data do not meet the Nyquist criterion for a
fully-sampled map and we have, therefore, made no attempt to generate
polarization maps at increased spatial sampling. All the polarization
data reported by \citet{hertzarchive} and used here maintain a
spatial resolution of $20\arcsec$. (Measured beam profiles are given here in
Figure~\ref{fig-beams} and in Figure~3 of \citet{hertz2}.)

The SCUBA camera and polarimeter is described in detail elsewhere
\citep{greaves2003,jenness2000, holland1999}. Briefly, the
$850\,\micron$ SCUBA-pol instrument consists of a single detector
array with 37 pixels arranged on a hexagonal grid with a $2\farcm3$
field-of-view; the polarization is measured by inserting a warm
wire-grid, modulated by a stepped HWP\@.  Fully-sampled polarimetric
and photometric maps are generated by ``jiggle-mapping'' which moves
the array by sub-pixel steps.  The data analysis involves combining
the individual jiggle maps and re-sampling them onto an output grid
with a $6\arcsec$ pitch.  Generation of fully-sampled maps in this
manner also alleviates the need to follow sky-rotation throughout the
night.  While the intrinsic SCUBA beam size at $850\,\micron$ is close
to the $14\arcsec$ diffraction limit of the JCMT, this map-making
process results in an effective beam-size closer to $20\arcsec$
(Figure~\ref{fig-beams}).


As noted above, the resultant spatial resolution for both Hertz at
$350\,\micron$ and SCUBA-pol at $850\,\micron$ is $\approx
20\arcsec$. We refer to this resolution as the effective beam-width,
which should not be confused with the diffraction limited resolution
of the respective telescopes.  Direct comparisons of the two beams
measured on Uranus are shown in Figure~\ref{fig-beams}. (The size of
Uranus at the time of each measurement was $3\farcs7$.)  Gaussian-fits
to these profiles yield $19\farcs5\pm0\farcs1$ for SCUBA-pol and
$19\arcsec\pm2\arcsec$ for Hertz; the reported statistical
uncertainties follow from a formal non-linear fit to the gaussian
profile. These beamsizes are consistent with the measurements reported
by \citet{hertz2} for Hertz ($20\arcsec\pm2\arcsec$) and
\citet{scubaflux} for SCUBA (a primary beam of $19\farcs5$).
Given this beam similarity we have made no correction for different
spatial resolutions when comparing these data sets.

The Hertz data are undersampled.  However, since the SCUBA data are
fully-sampled, there is sufficient information to estimate the SCUBA
intensity and polarization at the same sky locations of the Hertz data.
This is accomplished by reducing the SCUBA-pol data in the same
manner as presented in \citet{scubaarchive} but choosing to output the
data to grids and map-center locations which match the Hertz dataset.
Table~\ref{table1} lists the objects observed by both Hertz and
SCUBA-pol at 350 and 850 $\micron$; Table~2 (in the electronic version
only) gives a more complete list of the locations within each of the
clouds. 
Table~2 also includes data for all points at both wavelengths for the
polarization magnitudes and their ratio $P(850)/P(350)$, position
angles and their difference $\phi(850)-\phi(350)$, intensity values
and their ratio $F(850)/F(350)$, and uncertainties on all values. All
polarization magnitudes in Table~2 have been corrected for positive
bias (Section~\ref{sec-debias}). The best estimates of those values is
sometimes zero; as a result the ratio $P(850)/P(350)$ is reported as
\texttt{Nan} for cases in which $P(850)$$=$$P(350)$$=$0, \texttt{Inf}
for cases in which only $P(350)=0$, and equal to zero when only
$P(850)=0$.
 
%

\begin{deluxetable}{lcccccccccccc}
\tabletypesize{\small}
\rotate
\tablewidth{0pt}
\tablecaption{Object Summary and Polarization Ratios \label{tbl-object}}
\tablehead{
\colhead{} & \colhead{} & \multicolumn{3}{c}{data satisfying $P_\mathrm{c} > 0$} & \colhead{}
& \multicolumn{3}{c}{data satisfying $P_\mathrm{c} \geq 3\sigma_p$} & \colhead{}  
& \multicolumn{3}{c}{...also satisfying $\vert\Delta\phi\vert\tablenotemark{c} < 10\arcdeg$} \\
\cline{3-5} \cline{7-9} \cline{11-13} \\
\colhead{Source} & \colhead{Total\tablenotemark{a}}
& \colhead{Number\tablenotemark{b}} & \colhead{Median} & \colhead{MAD} &
& \colhead{Number\tablenotemark{b}} & \colhead{Median} & \colhead{MAD} &
& \colhead{Number\tablenotemark{b}} & \colhead{Median} & \colhead{MAD} 
}

\startdata
W3                            & \phn\phn 91 & \phn\phn 58 & 2.2 &  1.1 & &    \phn 14 &    2.5 &     0.8 & & \phn 4  & 2.0 &     0.3  \\
NGC\,1333                     & \phn    154 & \phn\phn 70 & 2.1 &  1.4 & & \phn\phn 3 &    3.5 &     1.8 & & \phn 1  & 6.5 & \nodata  \\ 
OMC-1                         & \phn    240 &    \phn 203 & 1.6 &  0.7 & &        136 &    1.6 &     0.6 & &     65  & 1.6 &     0.5  \\ 
OMC-2                         & \phn\phn 90 & \phn\phn 54 & 1.9 &  0.9 & &    \phn 10 &    2.4 &     0.6 & & \phn 2  & 2.6 &     0.3  \\ 
OMC-3                         & \phn\phn 99 & \phn\phn 82 & 2.3 &  1.1 & &    \phn 35 &    2.0 &     0.8 & &     15  & 1.6 &     0.6  \\ 
NGC\,2024                     & \phn    104 & \phn\phn 72 & 2.3 &  0.9 & &    \phn 19 &    2.8 &     0.6 & & \phn 5  & 2.2 &     0.2  \\ 
NGC\,2068 LBS\,10             & \phn\phn 62 & \phn\phn 45 & 1.7 &  0.6 & &    \phn 23 &    1.9 &     0.5 & & \phn 7  & 1.9 &     0.8  \\ 
NGC\,2068 LBS\,17             & \phn\phn 62 & \phn\phn 24 & 1.6 &  0.6 & & \phn\phn 0 &\nodata & \nodata & & \phn 0  & \nodata & \nodata  \\ 
NGC\,2071                     & \phn\phn 50 & \phn\phn 18 & 4.0 &  2.0 & & \phn\phn 1 &    2.4 & \nodata & & \phn 0  & \nodata & \nodata  \\ 
Mon\,R2                       & \phn\phn 76 & \phn\phn 58 & 3.2 &  1.4 & & \phn    22 &    2.8 &     0.9 & & \phn 1  & 1.0 & \nodata  \\ 
Mon\,OB1\,12\tablenotemark{d} & \phn\phn 66 & \phn\phn 47 & 3.7 &  2.1 & & \phn    11 &    4.8 &     1.6 & & \phn 1  & 9.8 & \nodata  \\ 
$\rho$\,Oph                   & \phn    100 & \phn\phn 76 & 1.7 &  0.8 & & \phn    25 &    1.3 &     0.4 & & \phn 7  & 1.3 &     0.5  \\ 
IRAS 16293$-$2422               & \phn\phn 63 & \phn\phn 18 & 5.2 &  1.9 & & \phn\phn 0 &\nodata & \nodata & & \phn 0  & \nodata & \nodata \\ 
NGC\,6334A                    & \phn\phn 59 & \phn\phn 37 & 2.7 &  1.6 & & \phn\phn 5 &    3.0 &     0.5 & & \phn 2  & 4.3 &     1.8  \\ 
W49\,A                        & \phn\phn 55 & \phn\phn 46 & 2.6 &  1.4 & & \phn    23 &    2.4 &     0.9 & & \phn 7  & 2.7 &     0.8  \\
W51\,A (G49.5-0.4)            & \phn    112 & \phn\phn 74 & 3.6 &  2.2 & & \phn    15 &    4.5 &     2.6 & & \phn 2  & 6.0 &     4.2  \\
DR\,21\tablenotemark{e}       & \phn    216 &    \phn 142 & 1.6 &  0.6 & & \phn    56 &    1.6 &     0.4 & &     22  & 1.6 &     0.4  \\
DR\,21 (Main)                 & \phn    100 & \phn\phn 72 & 1.7 &  0.6 & & \phn    36 &    1.6 &     0.5 & &     14  & 1.6 &     0.4 \\
\hline                                                            
All                           &        1699 &        1124 & 2.1 &  1.0 & &        398 &    1.9 &     0.7 & &    141  & 1.7 &     0.6 
\enddata

\tablecomments{The median polarization ratios ($P[850]/P[350]$), the
  median absolute deviation of their distribution
  (MAD; eq.~[\ref{eq-mad}]), and the number of data points in each sample
  are shown here. The column labeled ``$P_\mathrm{c} > 0$'' indicates
  regions where the measured data satisfy $P_\mathrm{m}/\sigma_p >
  \sqrt{2}$ at both wavelengths (see Section~\ref{sec-debias}).
  Columns labeled ``$P_\mathrm{c} \geq 3\sigma_p$'' indicate only data
  points satisfying that signal-to-noise criterion at both
  wavelengths; these values are calculated after applying the
  de-biasing technique discussed in Section~\ref{sec-debias}.  The
  columns labeled ``also $\vert\Delta\phi\vert < 10\arcdeg$'' satisfy
  both the latter $P/\sigma_p$ constraint as well as the additional
  $\Delta\phi$ criterion.\tablenotemark{c} Source coordinates
  can be found in Table~1 of \citet{hertzarchive}.}

\tablenotetext{a}{Total Number of points common to both the 350 and
  850 $\micron$ data sets.}

\tablenotetext{b}{Number of data points from the ``Total'' column
  satisfying the criteria above.}

\tablenotetext{c}{$\Delta\phi \equiv \phi(850) - \phi(350)$, where
  $\phi$ refers to the polarization position angle at the noted
  wavelength.}  \tablenotetext{d}{IRAS\,06382+0939}
\tablenotetext{e}{All data in DR\,21, including DR\,21 (Main).}

\label{table1}
\end{deluxetable}


%
%

\subsection{Positive Polarization Bias} \label{sec-debias}

By definition, the polarization amplitude is a positive-definite
quantity.  As a result, a noisy measurement of a truly
zero-polarization source will result in a mean positive
polarization. While there is no exact analytical method to correct for
this bias many authors use the formula: $P_\mathrm{c} \approx
(P^2_\mathrm{m} - \sigma_p^2 )^{1/2}$, where $P_\mathrm{c}$ is the
bias-corrected polarization, $P_\mathrm{m}$ is the measured
polarization, and $\sigma_p$ is the measured polarization
uncertainty. \citeauthor{plimits} (\citeyear{plimits}; also
\citealt{simmons85,quinn2012}) showed this was a good estimator when
$P_\mathrm{m} \gtrsim 3\sigma_p$ but that $P_\mathrm{c} = 0$ was a
better estimator if $P_\mathrm{m}/\sigma_p \leq \sqrt{2}$.
\citet{scubaarchive} applied the formula for high signal-to-noise to
all their $850\,\micron$ data while \citet{hertzarchive} applied no
corrections to their $350\,\micron$ data.  For the data presented in
this work we set $P_\mathrm{c} = 0$ in cases where
$P_\mathrm{m}/\sigma_p \leq \sqrt{2}$ and apply the above formula for
data with $P_\mathrm{m}/\sigma_p > \sqrt{2}$.  When computing
signal-to-noise cuts on $P/\sigma_p$ in Table~\ref{table1} and the
following sections we use the corrected values $P_c$ as described
above.  The conclusions drawn in Sections \ref{sec-compare} and
\ref{sec-discussion} use only the $P_\mathrm{c} \gtrsim 3\sigma_p$
sample and are therefore not effected by the fact that the correction
does not provide the best estimate within the regime $\sqrt{2} \leq
P/\sigma_p < 3$.

The corrections on $P$ have no effect on estimates of the polarization
position angle; that is in the sense that there is no bias in the
angle estimate as there is for the polarization amplitude. However,
there is necessarily an effect on the angle uncertainty (e.g.,
\citealt{clarke93}). For example, in cases where the
$P_\mathrm{m}/\sigma_p \leq \sqrt{2}$ the best estimate is
$P_\mathrm{c}=0$ and thus any angle measurement is meaningless. We
have made no attempt to estimate ``corrected'' angle uncertainties in
Table~2.  Such corrections are small for high signal-to-noise data
\citep{clarke93} and therefore they will have little effect on
our analysis and discussions below which use only data with
$P_\mathrm{c} \gtrsim 3\sigma_p$.

%
%
%
%
%

\section{Comparison} \label{sec-compare}

Maps comparing the $350\,\micron$ and $850\,\micron$ polarization data
in each region are shown in the Appendix.  Here we concentrate on
quantitative comparisons of the polarization magnitudes and angles
between the two different wavelengths.

The uncertainties on individual polarization measurements and on
combined quantities like their angle differences and ratios can be
quite large.  In the measurements below we concern ourselves with the
question of whether the observed distributions arise solely from
measurement uncertainties or also have significant contributions from
intrinsic variations across the clouds and/or locations within those
clouds.  To do this we define the reduced-$\chi^2$:
\begin{equation}
  \chi^2_r = \frac{1}{N-1} \sum_{i=1}^N \frac{(x_i - x_\mathrm{m})^2}{\sigma_i^2}
  \label{eq-chi2}
\end{equation}
where $x_\mathrm{m}$ is the median value of the samples $x_i$,
$\sigma_i$ is the measurement uncertainty on the quantity $x_i$, and
$N$ is the total number of data points in the sample.

\subsection{Polarization Angles} \label{sec-pangles}

\begin{figure*}
\centering
  \includegraphics[scale=0.5]{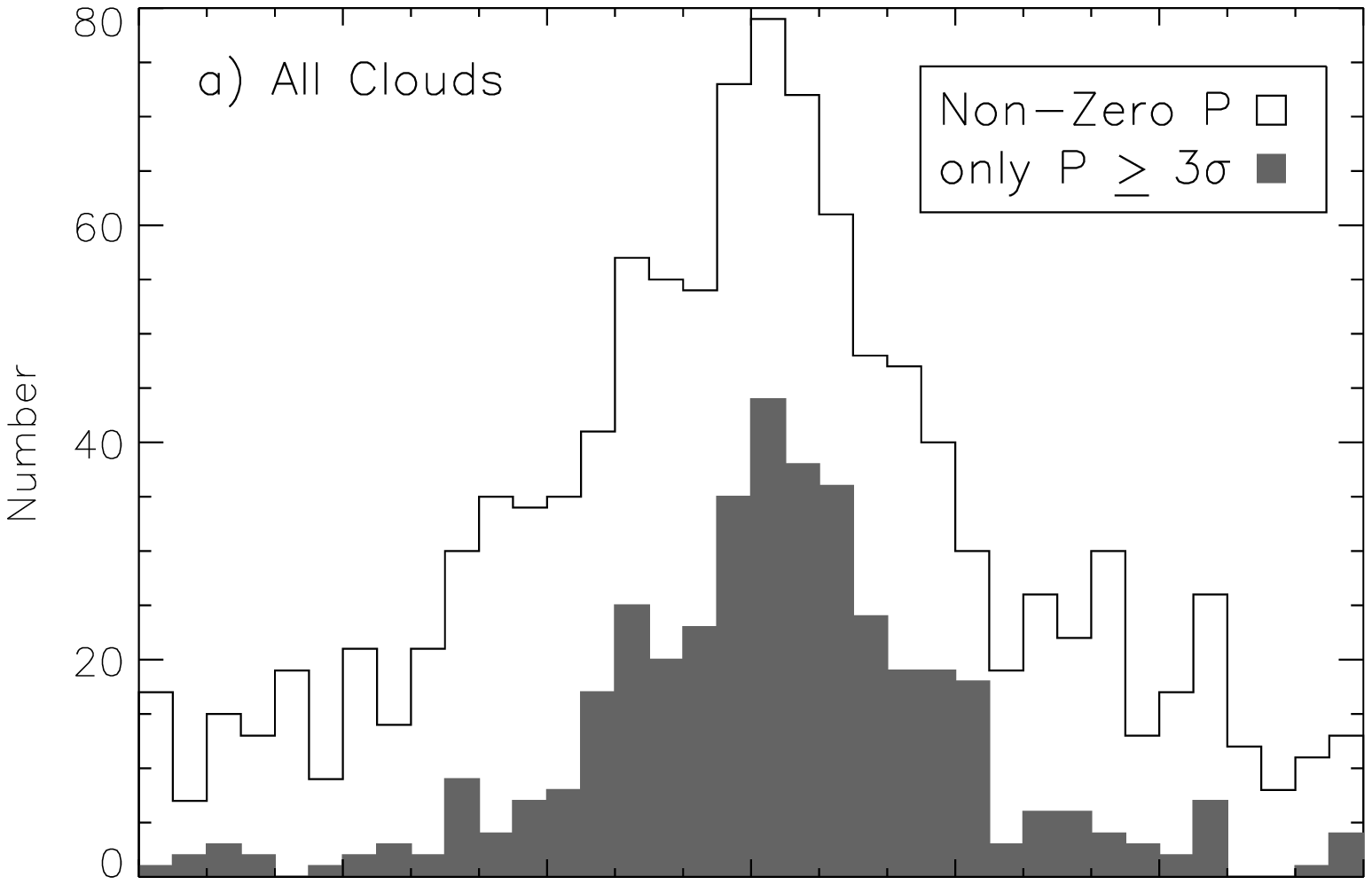}
  \includegraphics[scale=0.5]{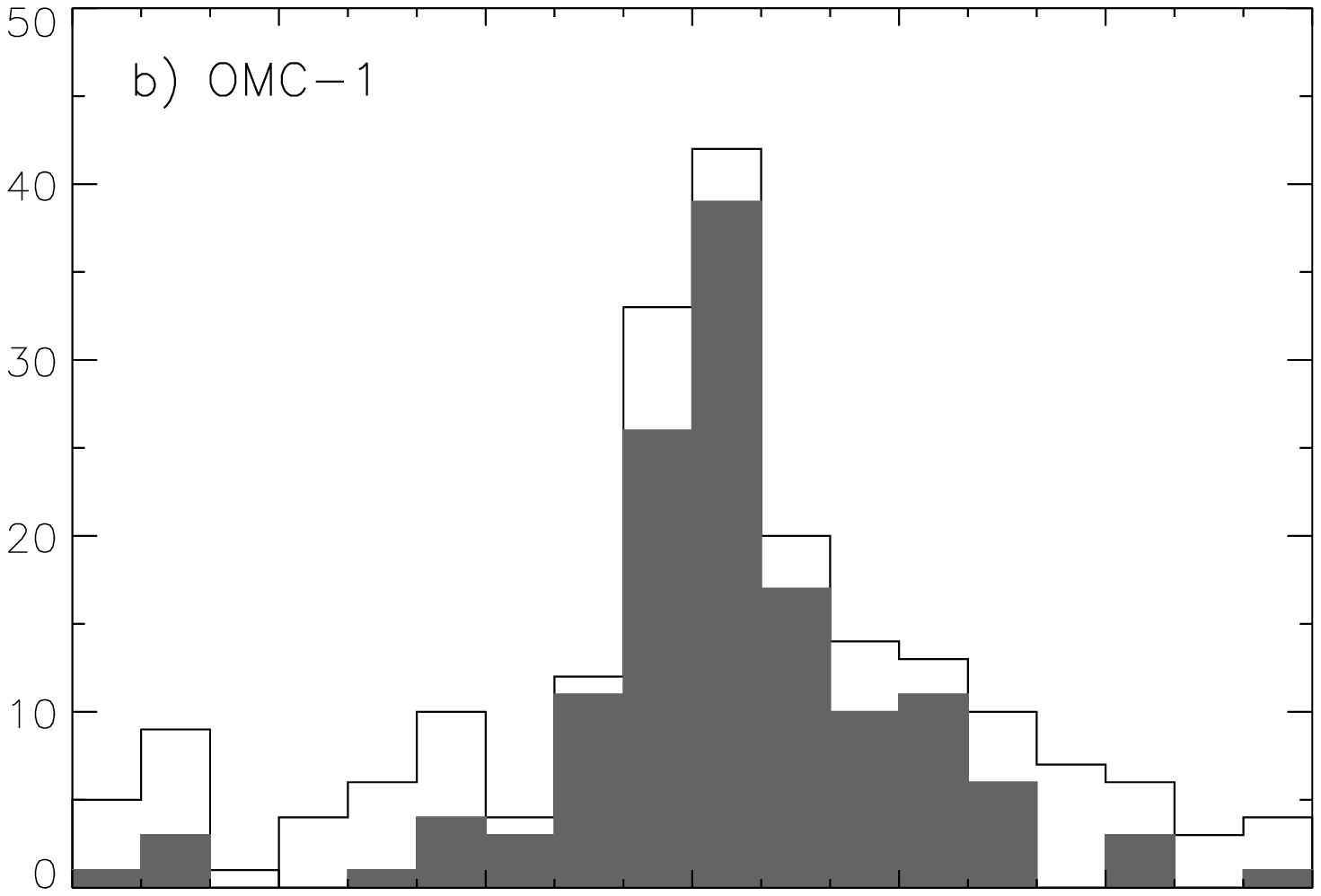}
  \includegraphics[scale=0.5]{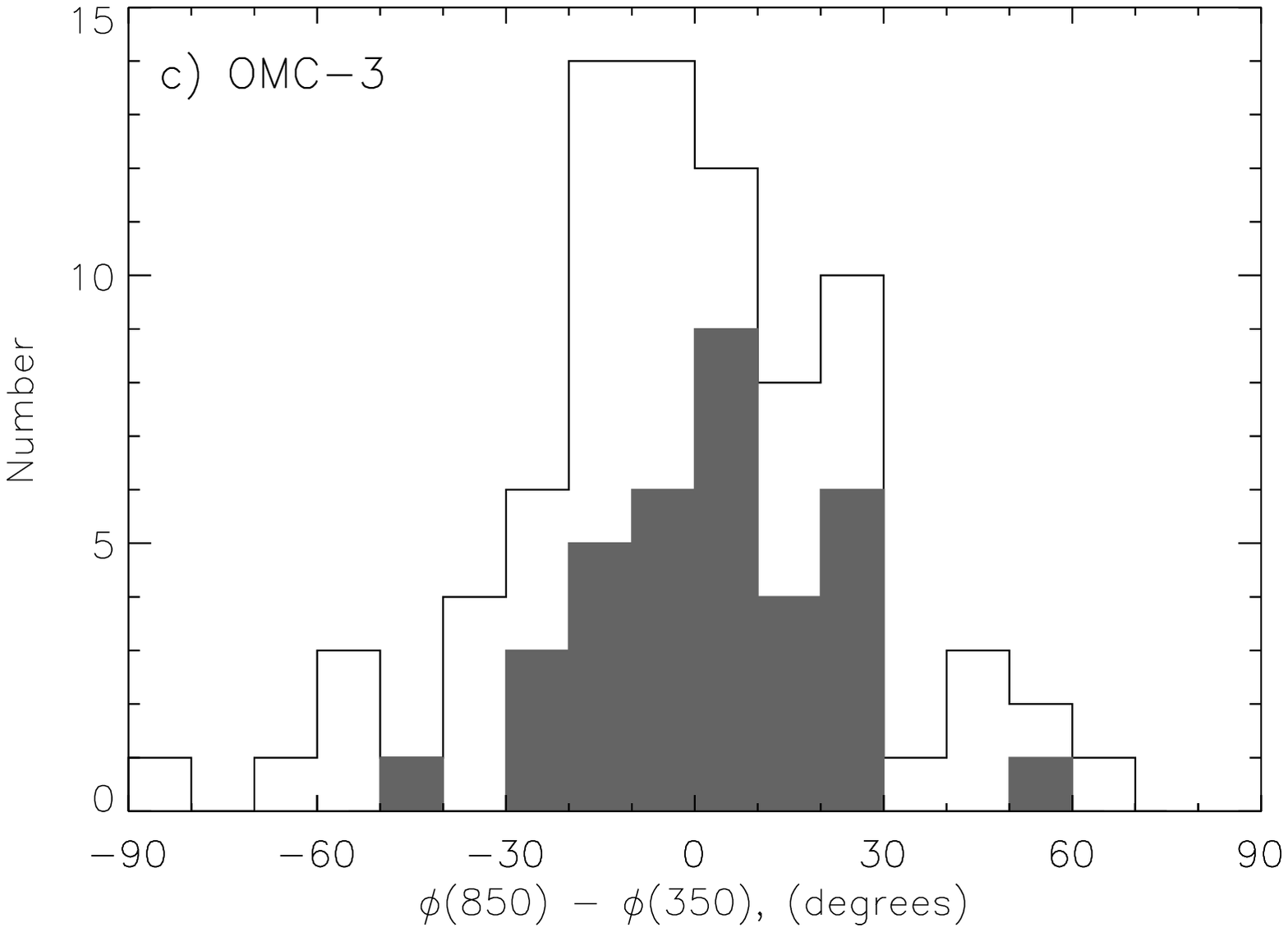}
  \includegraphics[scale=0.5]{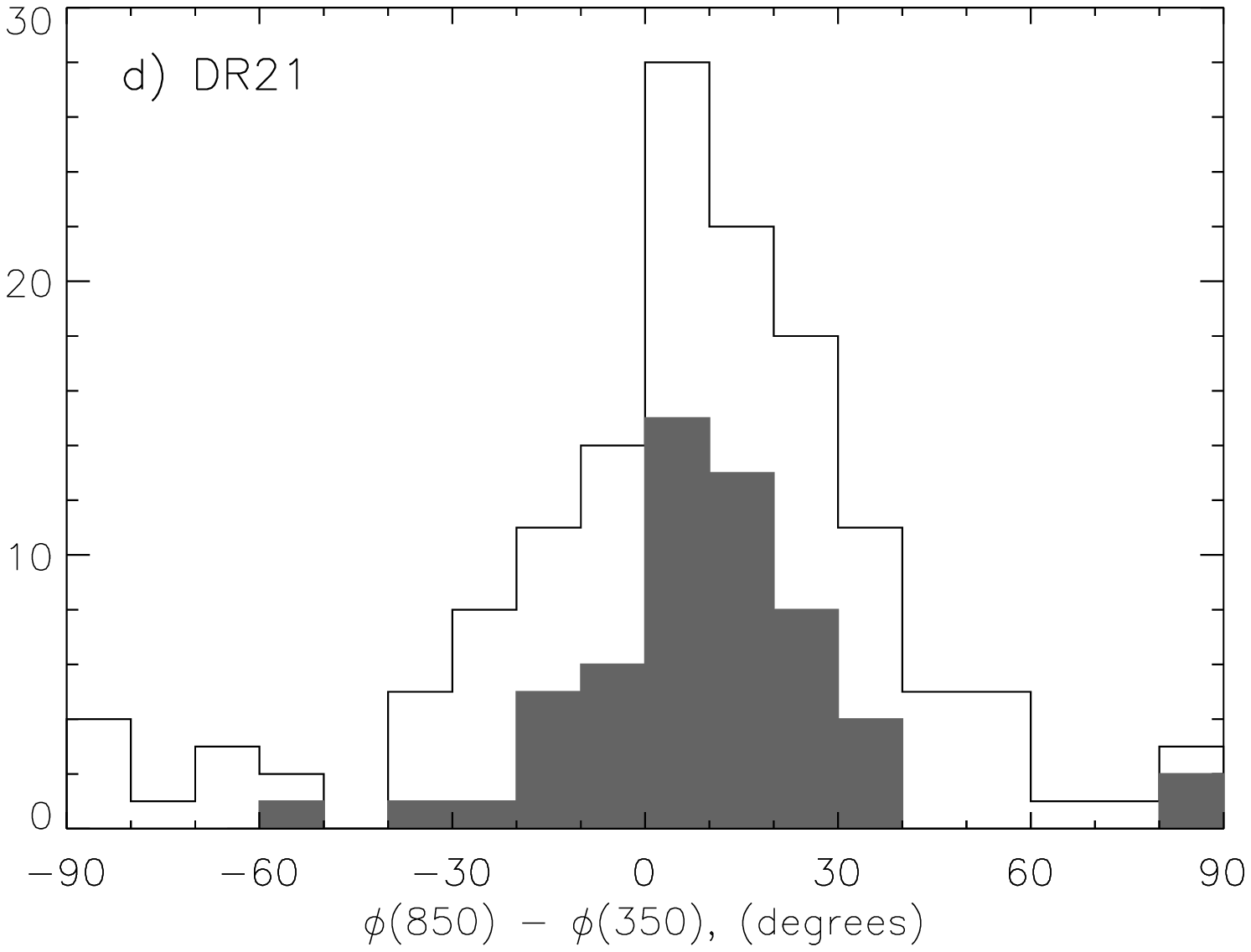}
  \caption{Histograms of the 850-to-350 $\micron$ polarization angle
    differences, $\phi_{850} - \phi_{350}$, for \emph{a}) all clouds
    in this work, \emph{b}) OMC-1, \emph{c}) OMC-3, and \emph{d})
    DR\,21. In each panel histograms are shown for all non-zero polarization data
    (open histograms) and only data satisfying $P \ge 3\sigma_p$ (gray
    histograms). The bin-widths for (\emph{a}) are $5\arcdeg$, all
    others are $10\arcdeg$. The total number of data points in each
    sample is given in Table~\ref{table1}.}
  \label{fig-anghist}
\end{figure*}

One of the most obvious aspects of the polarizations presented in the
maps is the agreement, or disagreement, of the position angles between
the two wavelengths. Figure~\ref{fig-anghist} compares these angles
across the entire data-set and within three clouds with the largest
number of data points. All the distributions peak near angle
differences of zero degrees, that is $\Delta\phi \equiv \phi(850) -
\phi(350) = 0\arcdeg$.  For all non-zero polarization data (i.e., data
where both $P(350)>0$ and $P(850)>0$) the median angle difference is
$1\arcdeg$ with a standard deviation of $39\arcdeg$.  This agreement
is stronger (i.e., has a smaller deviation) if we limit ourselves to
only points with $P \ge 3\sigma_p$ (398 points in the entire data
set). In this case the median angle difference is $4\arcdeg$ with a
standard deviation of $28\arcdeg$.

To rule-out the possibility that the width of the angle distribution
is strongly dependent on the measurement uncertainties we calculate the
$\chi_r^2$ value of the angle differences following equation~(\ref{eq-chi2}).
If most data points were consistent with the median angle difference
within their uncertainties then we would expect $\chi_r^2\sim1$
(especially for $N-1=397$ degrees-of-freedom for the entire $3\sigma$
data-set).  However, for the complete data set with $P \ge 3\sigma_p$
we find $\chi_r^2 = 14$\@.
From this we conclude the distribution's width is intrinsic and not a
result solely of data uncertainties.  

Most of the clouds in our sample have insufficient data to perform
this analysis separately on individual clouds. Exceptions to this
point are OMC-1, OMC-3, and DR\,21, whose angle distributions are also
shown in Figure~\ref{fig-anghist}. For data satisfying the
$P\ge3\sigma_p$ criterion in those three clouds the median angle
differences and standard deviations are $3\arcdeg\pm25\arcdeg$,
$4\arcdeg\pm19\arcdeg$, and $10\arcdeg\pm22\arcdeg$, with $\chi^2_r$
values of 18, 5, and 7, respectively. Therefore, as was observed for
the entire data set above, the width of the angle-difference
distribution ($\Delta\phi$) in these individual clouds is real, in
the sense that they are not a result solely of the measurement
uncertainties.

Lastly, we should note that the position angle rotations observed in
these clouds are unlikely to be the result of Faraday rotation. In
typical interstellar cloud conditions, at these wavelengths, Faraday
rotation is generally much smaller than that observed here (e.g.,
\citealt{w3}; \citealt*{mwf01}).

\subsection{Polarization Ratio} \label{sec-pratio}

%

\begin{figure*}
  \centering
  \includegraphics[scale=0.5]{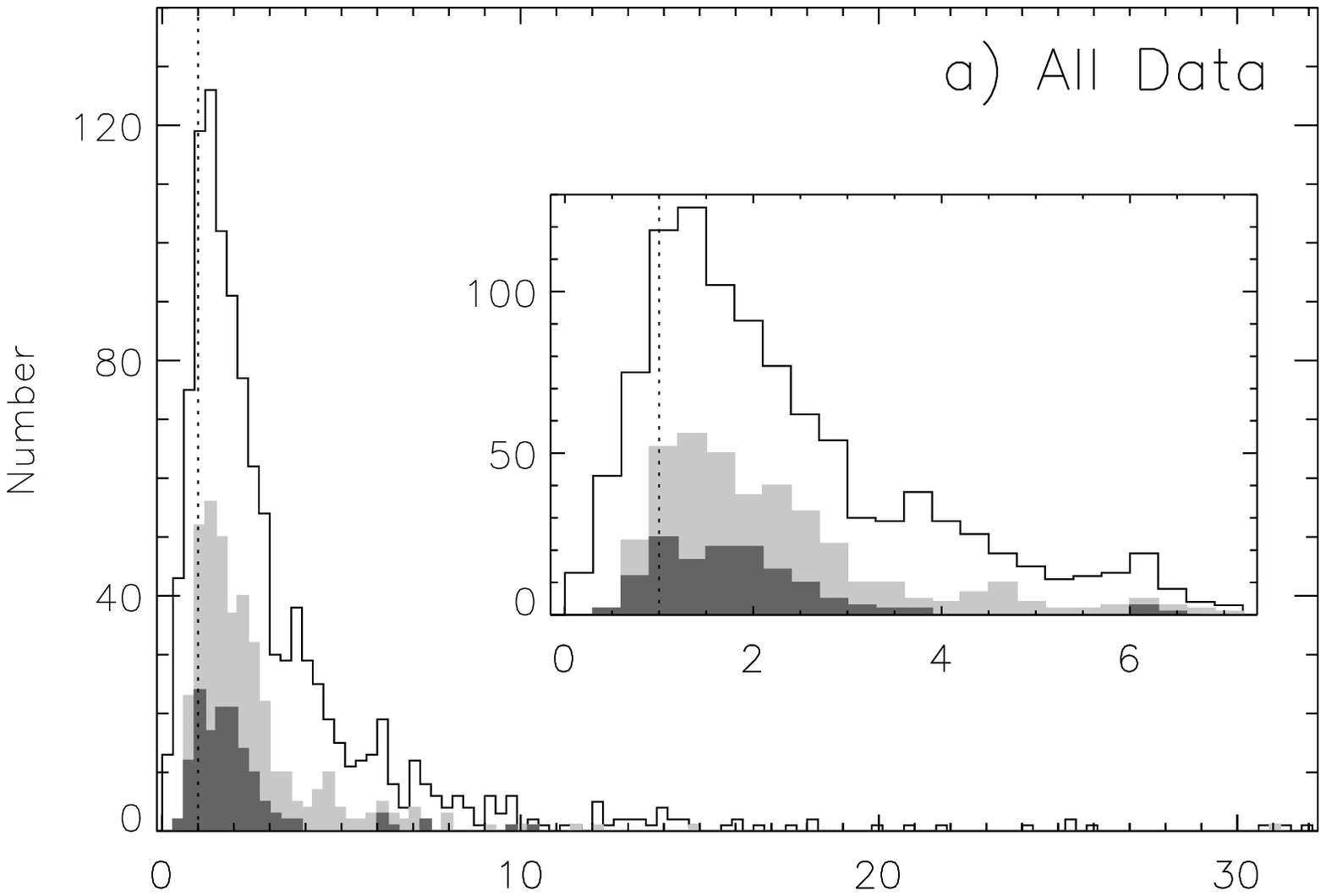}
  \includegraphics[scale=0.5]{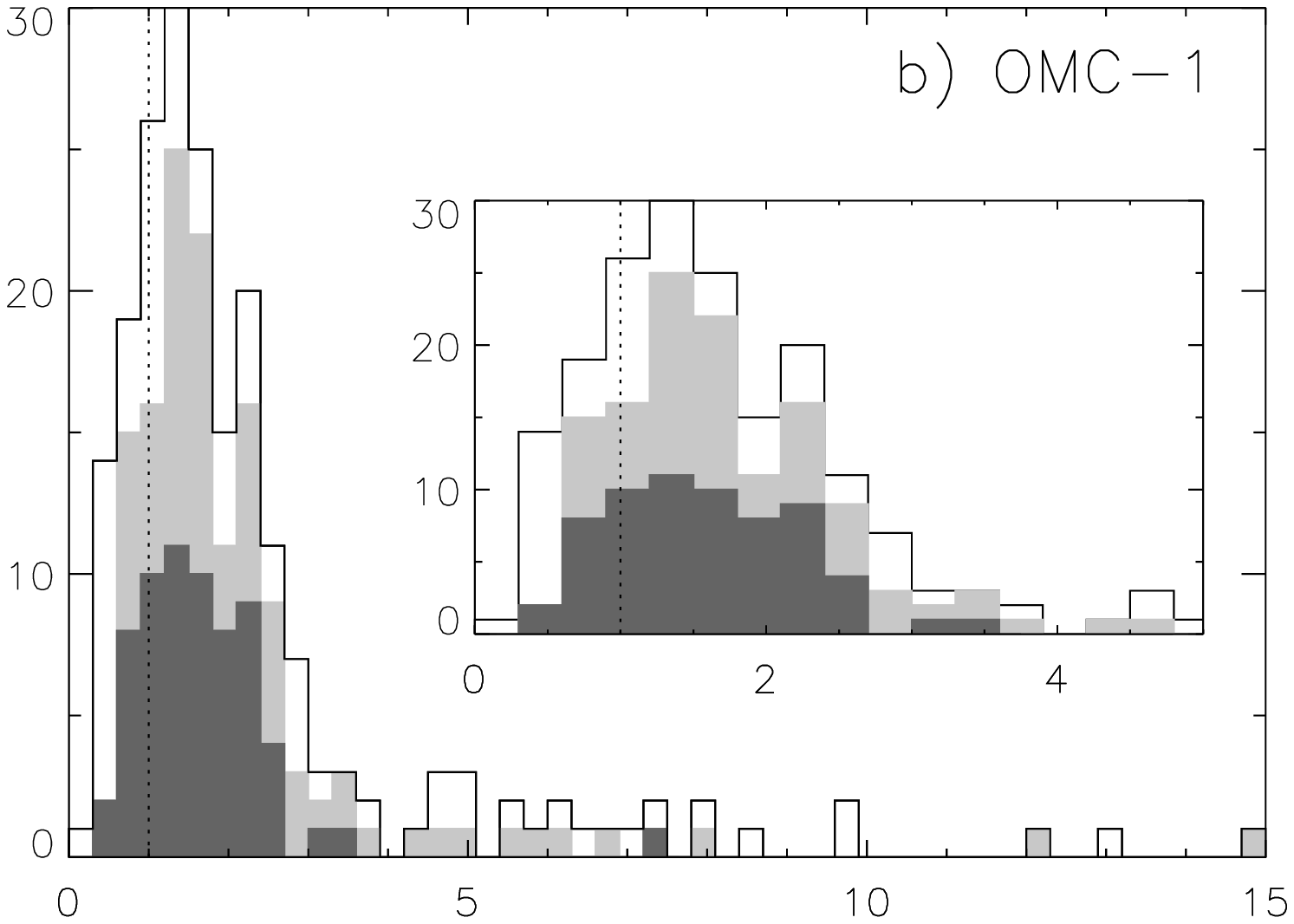}
  \includegraphics[scale=0.5]{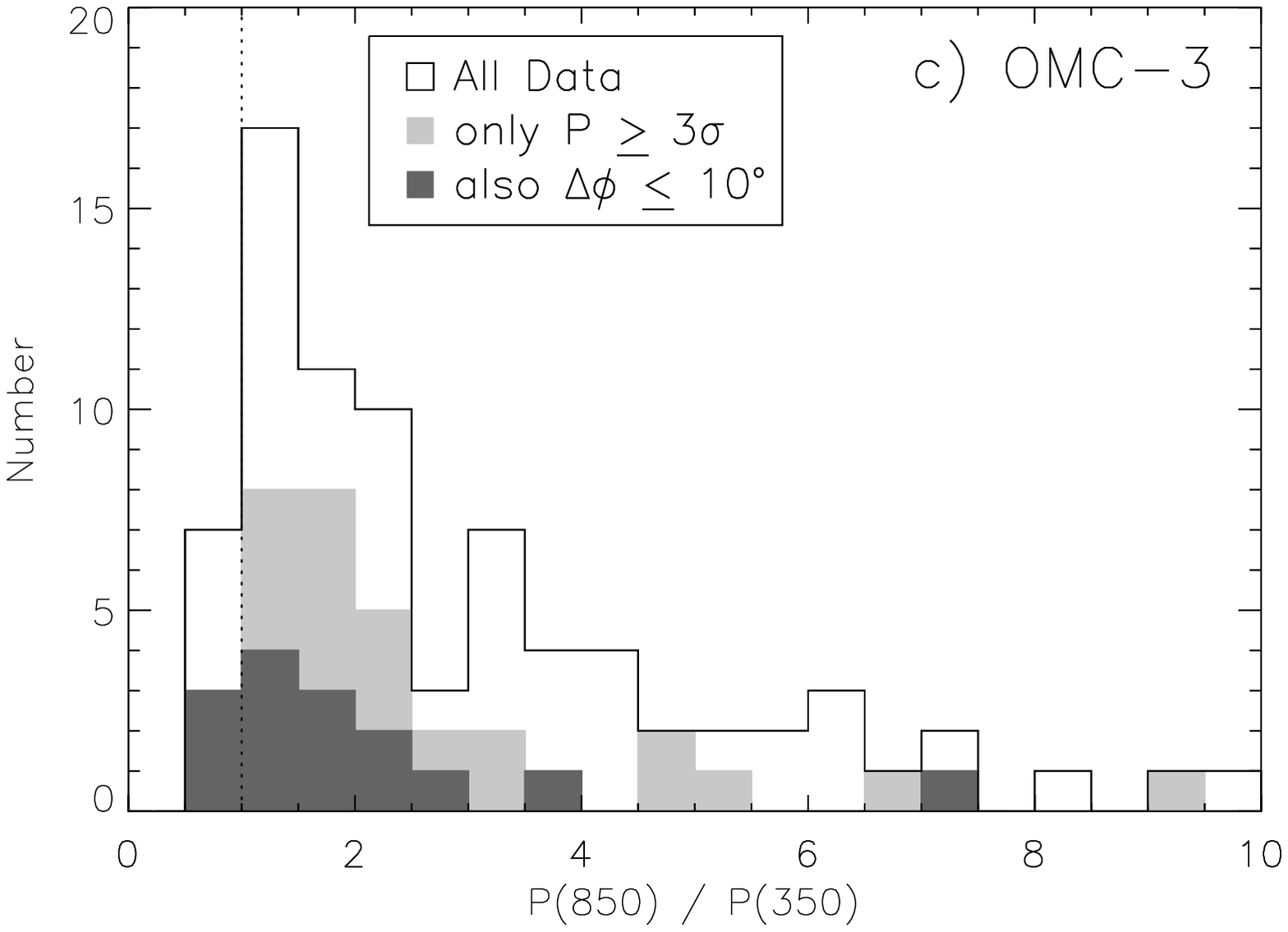}
  \includegraphics[scale=0.5]{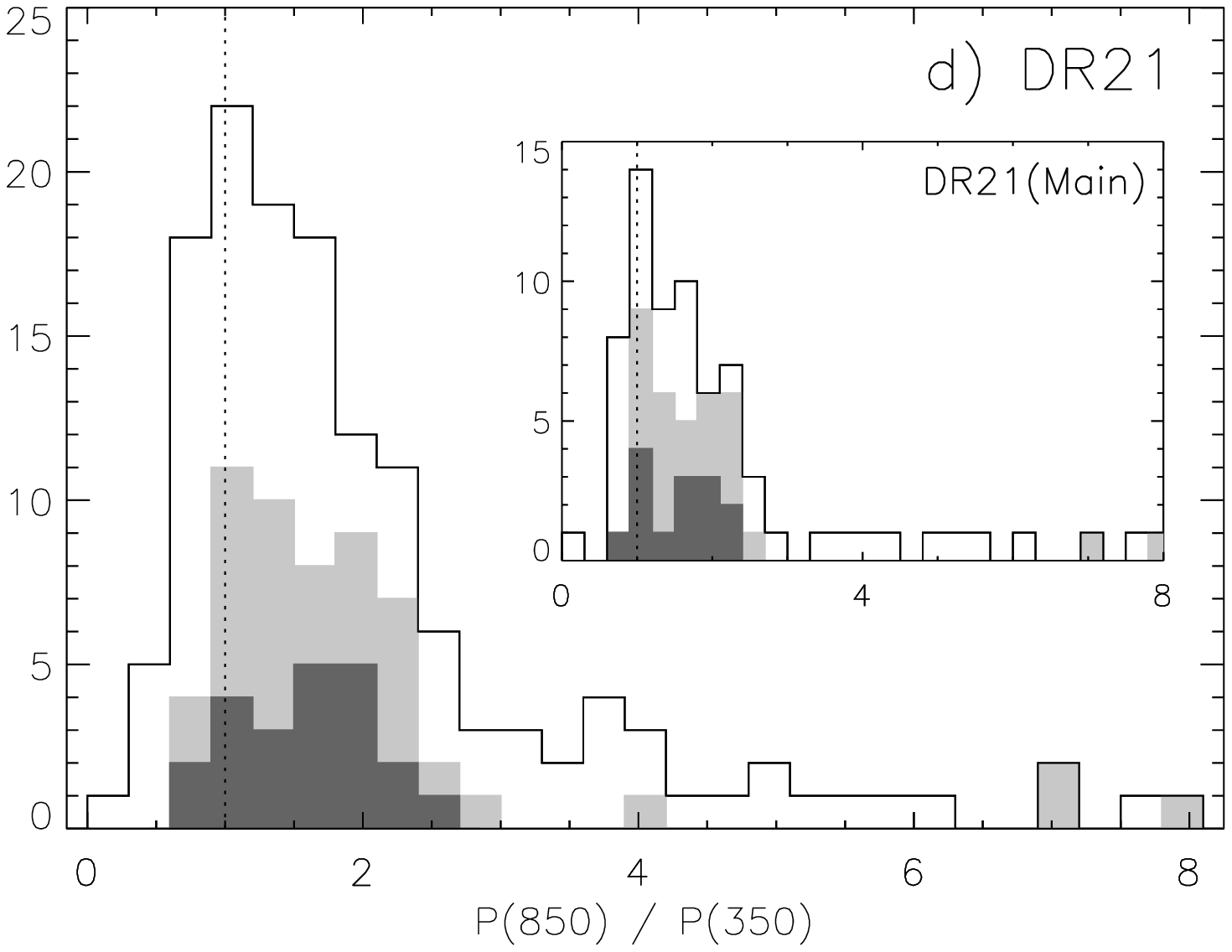}
  \caption{Histograms of the 850-to-350 $\micron$ polarization ratio
    for (\emph{a}) all data in this work, (\emph{b}) OMC-1, (\emph{c})
    OMC-3, and (\emph{d}) DR\,21. Each plot contains different
    cuts on the data as shown in the legend of (\emph{c}): all data
    (open histogram), only $P \ge 3\sigma_p$ (light gray), $P \ge
    3\sigma_p$ and $|\Delta\phi\vert \le 10\arcdeg$ (dark gray). For 
    reference vertical dotted lines are drawn at $P(850)/P(350)=1$. For
    display purposes only, the insets in (\emph{a}) and (\emph{b})
    magnify the lower-end of the distributions where the histograms
    peak.  Note the larger plot in (\emph{d}) contains all data in DR21 
    while the inset shows only data in DR21(Main). The bin-widths for 
    (\emph{c}) are 0.5, all others are 0.3.
    The total number of data points in each sample is given in
    Table~\ref{table1}.}
  \label{fig-polhist}
\end{figure*}

Table \ref{table1} lists the total number of locations where
measurements were made at both wavelengths.  For the best comparisons
we typically choose to include only data satisfying the
signal-to-noise criterion $P \ge 3\sigma_p$; this criterion is applied
after the de-biasing correction discussed in Section \ref{sec-debias}.
Figure~\ref{fig-polhist}\emph{a} shows the distribution of these
points. This figure shows all data satisfying the $3\sigma_p$
criterion, including outliers as high as $P(850)/P(350)=31$; the inset
concentrates on the main peak.

As can be seen in Figure~\ref{fig-polhist} the distributions are non-normal
in nature and often contain outliers away from the main peaks.  
Therefore, unlike the polarization angle distributions in
Section~\ref{sec-pangles}, none of these distributions are well
characterized by a simple sample mean or a sample standard
deviation. As robust descriptions of the distributions' central
tendencies and width we use the samples' medians and median absolute
deviations (MAD\@).  The MAD of a set of measurements $x$ is defined
as the median value of the residuals, where the residuals are also
calculated with respect to the sample median; that is
\begin{equation}
\mathrm{MAD}\equiv \mathrm{median}\left(\vert x - x_\mathrm{m} \vert\right).
\label{eq-mad}
\end{equation}
where $x_\mathrm{m}$ is the median value of the measurements $x$. For
a normal distribution the MAD is significantly smaller than its
standard deviation ($\sigma$) with an expectation value of
$\sigma/1.48$. However, given the small-number of comparison points in
many of the clouds in Table~\ref{table1} and that the
$P(850)/P(350)$-distribution is not expected to be symmetric
we report only the MADs there.


The complete $3\sigma$ data set contains 398 points
with a median $P(850)/P(350)$ value of 1.9, MAD=0.7, and
$\chi^2_r$$=$$9$.  The $\chi_r^2$-value implies that the
distribution's width is intrinsic and not a result solely of data
uncertainties.  We reach the same conclusion examining the
distributions for OMC-1, OMC-3, and DR21. (Medians are shown in
Table~\ref{table1}, $\chi_r^2 = $ 12, 5, and 3.)

We note that many of the peaks in the distributions of
Figure~\ref{fig-polhist} are clearly different from the medians listed
in Table~\ref{table1}, this is mostly driven by some large outliers in
the distribution.  An alternate estimate of this peak is the value
which minimizes the MAD\@.
For the entire $3\sigma$ data-set 
in Figure~\ref{fig-polhist}\emph{a} this alternative peak estimate is
1.5 with MAD=0.6\@.  If the median in equation~(\ref{eq-chi2}) is
replaced with this peak then $\chi^2_r = 6$. These peaks, new MAD's,
and $\chi^2_r$ of OMC-1, OMC-3, and DR21 are given in
Table~\ref{table3}. Using these peaks yields $\chi^2_r>1$ for all
three clouds and, therefore, does not change the conclusion that the
distributions' widths are intrinsic and not a result solely of data
uncertainties.  Also, the difference between the median and the peak
is less than the MADs in all cases (i.e., the whole data-set and the
three specific clouds); none of the discussion in
Section~\ref{sec-discussion} relies strongly on these precise values.

%
\addtocounter{table}{1}
\begin{deluxetable}{lccccccc}
\tabletypesize{\small}
\tablewidth{0pt}
\tablecaption{Polarization Ratio Distributions \label{tbl-distrib}}
\tablehead{
\colhead{}
& \multicolumn{3}{c}{data satisfying $P \geq 3\sigma_p$} & \colhead{}  
& \multicolumn{3}{c}{...also satisfying $\vert\Delta\phi\vert < 10\arcdeg$} \\
\cline{2-4} \cline{6-8} \\
\colhead{Source}
& \colhead{Peak} & \colhead{MAD} & \colhead{$\chi^2_r$}
& \colhead{} & \colhead{Peak} & \colhead{MAD} & \colhead{$\chi^2_r$}
}

\startdata
OMC-1        &    1.3 &   0.5 &    13.4 &  & 1.4 & 0.5 &    20.8  \\
OMC-3        &    1.4 &   0.5 & \phn2.9 &  & 1.2 & 0.3 & \phn3.7  \\
DR\,21       &    1.4 &   0.4 & \phn2.5 &  & 1.8 & 0.3 & \phn5.2  \\
DR\,21(Main) &    1.3 &   0.3 & \phn2.3 &  & 1.8 & 0.3 & \phn6.6  \\
All          &    1.5 &   0.6 & \phn6.0 &  & 1.4 & 0.5 &    10.9
\enddata

\tablecomments{The peak value and the median absolute deviation (MAD)
  of the polarization ratio ($P[850]/P[350]$) distributions which
  minimize the MAD (eq.~[\ref{eq-mad}]). Also shown are $\chi^2_r$
  values as calculated from equation (\ref{eq-chi2}); see
  Section~\ref{sec-pratio}. The columns labeled ``$P \geq 3\sigma_p$''
  and ``also $\vert\Delta\phi\vert < 10\arcdeg$'' are defined as in
  Table~\ref{tbl-object}.}

\label{table3}
\end{deluxetable}


In using the polarization ratio to study grain alignment (e.g.,
Section~\ref{sec-grains}) we want to ensure that data at both
wavelengths are sampling the same regions of the cloud, both along the
line-of-sight (LOS) and across the plane-of-the-sky (POS)\@.  The
latter criterion is simply met due to the fortuitous ability to
beam-match the 350 and 850 $\micron$ data. If the emission sources are
the same for radiation at both 350 and 850 $\micron$ then the former
criterion would also be met. Meeting this criterion is more difficult,
but we try to limit its effect by choosing data with little-to-no
position angle rotation between the two wavelengths. For this reason
our analysis is often limited to data points where
$\vert\Delta\phi\vert \le 10\arcdeg$. Note that $\sigma_\phi \le
10\arcdeg$ corresponds to data points with $P \gtrsim 3\sigma_p$.

To understand this particular data-cut, consider the case where the
magnetic field changes its orientation along the line-of-sight (and
within the cloud depths sampled by at least one wavelength). This may
result in a wavelength-dependent change in both the polarization
position angle (which follows the change of the field's projected
orientation) and the polarization level (due to the field's changing
inclination angle). In interpreting the polarization spectrum in terms
of grain alignment physics, we wish to eliminate the changing
inclination angle as the cause of any change in the polarization level
(which can also result from other grain/cloud properties; see
Section~\ref{sec-grains}). Since a changing field orientation must
occur for any data with a wavelength-dependent angle,
we can eliminate regions where this occurs by removing such data.
While this does not ensure that points without wavelength-dependent
angles arise from a single source it is unlikely, in the statistical
sense, that the LOS field angles can change for many points in our
large sample without an accompanying POS rotation.

Table~\ref{table1} shows the total number of data points satisfying
both the $P/\sigma_p \ge 3$ and $\vert\Delta\phi\vert \le 10\arcdeg$
data-cuts in each cloud, along with their median polarization ratios
and median absolute deviations. After making this second data cut the number
of surviving data points drops to 141, only 35\% of the
$3\sigma$ dataset.  
The resulting median ratio is $P(850)/P(350) = 1.7$ with MAD=0.6 and
$\chi_r^2 = 11$ (see Figure~\ref{fig-polhist}\emph{a}).  The
$\chi_r^2$ values for OMC-1, OMC-3, and DR21 are also large using this
data cut ($\chi_r^2 = $ 19, 5, and 3, respectively), again implying
that the distributions' widths are not a result solely of data
uncertainties. There are not as many outliers in the distributions
after the $\vert\Delta\phi\vert$-cut as their were in the
$3\sigma$-only cut.  These do have some effect on the measured MADs
(as discussed earlier in this section); Table~\ref{table3} reports the
distribution peaks, MADs, and revised values of $\chi_r^2$ for the
$\vert\Delta\phi\vert$-cut in the case where we have minimized the
MADs. These small changes in the distribution widths
still result in values of $\chi_r^2 \gtrsim 1$ meaning that our
conclusion, that the widths are intrinsic and not a result solely of
data uncertainties, still holds.

\section{Discussion} \label{sec-discussion}

\subsection{The Polarization Spectrum}



\begin{figure}
  \plotone{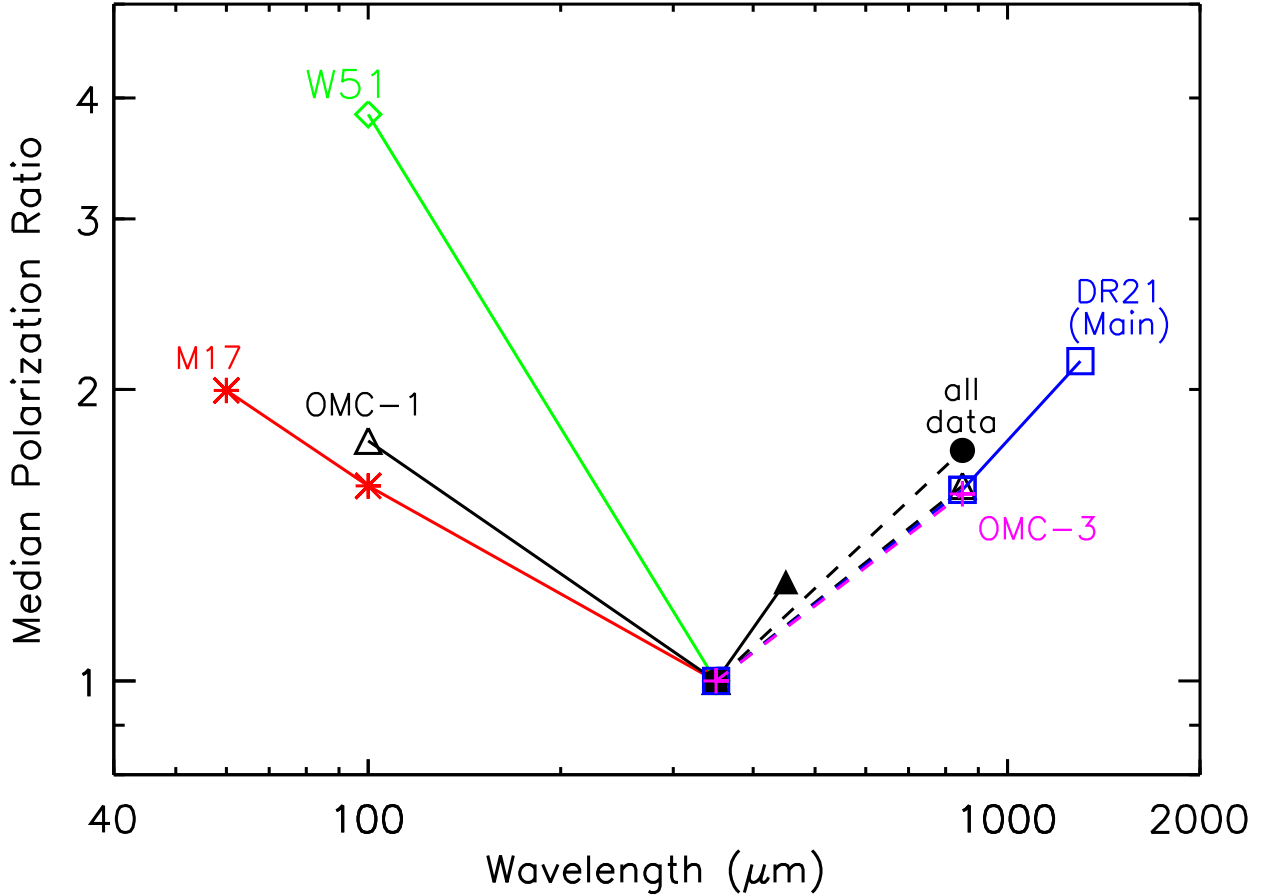}
  \caption{Polarization spectrum in several molecular clouds,
    normalized at $350\,\micron$.  The $P(850)/P(350)$ data in this
    work are shown as dotted lines; note that OMC-1, OMC-3, and
    DR21(Main) have identical medians (see Table~\ref{table1}). The
    solid circle represents the median of all data in this work.  All
    data in this plot, including that shown here for the first time,
    meet the $3\sigma_p$ and $\Delta\phi$ criteria described in the
    text.  Data at $\lambda<350\,\micron$ and that for DR21(Main) at
    $P(1300)/P(350)$ are from \citet{mythesis}.  All data used at
    $350\,\micron$ are from Hertz \citep{hertzarchive} with the
    exception of the OMC-1 point comparing 350 and 450~$\micron$
    (solid triangle) which is from SHARP \citep{omc1sharp}.  For
    clarity no error bars are shown here, but see
    Figure~\ref{fig-polhist} for the distributions. [\emph{A color
      version of this figure is available in the electronic
      version.}]}
  \label{fig-pspec}
\end{figure}

Figure~\ref{fig-pspec} shows an updated version of the polarization
spectrum from \citet{mythesis} and \citet{omc1sharp}.  All the data in
this figure satisfy the criteria $P\ge 3\sigma_p$ and
$\vert\Delta\phi\vert \le 10\arcdeg$. 
Here we plot the median value of DR21(Main), rather than DR21, as the
data at other wavelengths ($1300\,\micron$) only cover that
region of the cloud. While all the data in Figure~\ref{fig-pspec} show
$P(850) > P(350)$ we would draw the reader's attention to the range of
ratios plotted in the distributions of Figure~\ref{fig-polhist}. For
example, the median ratio for all clouds in this work is plotted at
$P(850) / P(350) = 1.7$ but has a relatively large MAD (0.6).

The previous work comparing Hertz and SCUBA-pol performed by
\citet{mythesis} and \citet{omc1sharp} were based on slightly
different data sets than we use here.  First, the earlier results were
obtained using the data prior to the systematic re-analyses performed
by \citeauthor{hertzarchive} (\citeyear{hertzarchive}; see also
\citealt{kirby05}) and \citet{scubaarchive}. Second, those results
attempted to match the Hertz and SCUBA beams by smoothing the SCUBA
data to match Hertz's presumably larger beam size and re-sampled at a
rate of 5 arcseconds per pixel.  Given the pointing accuracies of
Hertz (4\arcsec--6\arcsec; \citealt{hertz2}) and SCUBA
(2\arcsec)\footnote{\url{http://www.jach.hawaii.edu/JCMT/telescope/pointing/pointing\_history.html}}
the $5\arcsec$ re-sampling was not unreasonable. However, as shown in
Figure~\ref{fig-beams}, the smoothing step was likely unnecessary.
Despite these analysis differences the median results are in good
agreement.  For data satisfying $P\ge 3\sigma_p$ and
$\vert\Delta\phi\vert \le 10\arcdeg$ the previous work found
$P(850)/P(350)$ medians and standard deviations of $1.4\pm0.6$ and
$1.7\pm2.7$ for OMC-3 and DR\,21(Main), respectively \citep{mythesis}.
Here we find medians and MADs of $1.6\pm0.6$ and $1.6\pm0.4$ for those
two clouds. 
The large difference in the DR21(Main) standard deviations (=1.5 for
the current work) likely lies in the fact that the earlier $850/350$
data comparison used a different set of DR21 data.
The analysis by \citet{scubaarchive} used additional observations not
available at the time of \citet{mythesis} and the analysis
resulted in better rejection of noisy data.

For W51, \citet{omc1sharp} found $1.8\pm2.4$ (median and standard
deviation). This is consistent with the W51 results shown in
Table~\ref{tbl-object}, $6.0\pm4.2$ (median and MAD), but only because
of the large deviations. These data are not plotted in
Figure~\ref{fig-pspec} as only two data points survive the data cuts.

The $P(850)/P(350)$ ratio for OMC-1 is calculated in this work for the
first time.  Figure~\ref{fig-pspec} also includes $P(450)/P(350)$
(solid triangle; \citealt{omc1sharp}) and $P(100)/P(350)$ (open
triangle; \citealt{mythesis}) data points.  The OMC-1 cloud is thus
the only cloud in our data set which includes data both above and
below the $350\,\micron$ minimum in the spectrum.

\subsection{Grain Alignment} \label{sec-grains}

The original work by \citet{pspec} made it clear that the simplest
model, isothermal dust populations all with the same polarization
and/or alignment properties, can not reproduce a spectrum like that in
Figure~\ref{fig-pspec}.  In fact such a model yields a polarization
spectrum independent of wavelength beyond 50\,\micron.
In order to explain such a spectrum we consider that a number of
physical mechanisms are responsible for the absolute polarization
level observed in dust emission.  Foremost among these are the
efficiency with which dust grains become aligned with magnetic fields
and variations in the inclination of that field to the
line-of-sight. Ideally, our $\vert\Delta\phi\vert\le 10\arcdeg$ data
cut (Section~\ref{sec-pratio}) has eliminated the field inclination as
a variable in the observed spectrum, leaving alignment efficiency as
the key variable.

In order to generate wavelength-dependent polarization spectra the
alignment efficiencies must be correlated (or anti-correlated) with
changes in the grains' emission. In order to explain spectra like
those in Figure~\ref{fig-pspec} \cite{pspec} considered simple
emission laws of the form $F(\nu)\propto\nu^\beta B_\nu(T)$, where
$\nu$ is the observed frequency, $\beta$ the spectral index, and
$B_\nu(T)$ is the Planck function at temperature $T$\@.  For such
models the required change in emission can take the form of
differences in temperature, differences in the spectral index, or a
combination of the two (see also \citealt{mythesis,paris}).  Below we
discuss two physical models of the ISM and molecular clouds, both of
which result in grain populations with the different alignment
properties and temperatures/spectral-indices which lead to
wavelength-dependent polarization spectra.


Theoretical models of grain alignment have a long history (see reviews
by \citealt{alexreview1,alexreview2,rhh88}) with detailed
observational tests possible only very recently (e.g.,
\citealt*{lgm97,matsumura2009};
\citealt{andersson2010,andersson2011,matsumura2011}).  In one of the
most recent models, that of ``radiative alignment torques'' (RAT;
\citealt{cho05,lazarian2007,hoang2008,hoang2009,hoang2009a})
stellar and interstellar photons provide the necessary torques to
align the spin-axes of dust grains parallel to the local magnetic
field.  \citet{bethell07} simulated a molecular cloud containing
aspherical graphite and silicate grains with a typical interstellar
grain-size distribution (i.e., \citealt*{mrn77}) and radii of
0.005--0.5~\micron. The RAT model results in alignment only of grain
sizes larger than some cut-off, the exact value of which is dependent
on properties like the gas density and radiation field and can vary
throughout the simulated cloud \citep{cho05}. As the larger grains are
more efficient emitters they reach cooler temperatures than the
smaller grains in equilibrium. This yields an anti-correlation between
grain temperature and alignment; the small warm grains are unaligned
while the large cool grains are aligned.  The resulting polarization
spectrum rises from 100 to 400~$\micron$, with little variation
at longer wavelengths.  For the wavelengths of interest here they find
$P(850)/P(350) \sim 1.0$--1.1.

\citet{draine09} also present models composed of aspherical silicate
grains and spherical graphite grains (here we discuss only their Model
numbers 1 and 3). The grain-size distribution and the relative
silicate-graphite mix are constrained by the observed interstellar
extinction. Additionally, rather than model any physical alignment
mechanism (such as RAT), their grain alignment is empirically
constrained by the typical interstellar polarization spectrum spanning
near-optical wavelengths (i.e., the ``Serkowski law'').
The result is similar to the work of \citet{bethell07} in the sense
that larger grains are both cooler and better aligned than the smaller
grains and produces a steep polarization spectrum in the 40 --
400~$\micron$ range.  However, the \citet{draine09} spectrum continues
to rise beyond $400\,\micron$ such that $P(850)/P(350) \sim 1.2$--1.3.
This long-wavelength behavior is a combination of a) the cooler
silicate grains being aligned, whereas the warmer spherical graphite
grains are not, and b) the shallower spectral indices of silicates
compared to graphites.


The median $P(850)/P(350)$ values presented in this work
(Fig.~\ref{fig-pspec} and Table~\ref{table1}) are clearly steeper than
the model estimates just discussed ($\sim1.7$ for the ``all clouds,
$P$$>$$3\sigma$, $\vert\Delta\phi\vert$$<$$10\arcdeg$''
sample). However, the data distributions are large and, therefore,
cannot strongly rule-out either model.  Additionally, the models are
calculated using physical conditions and constraints which likely do
not prevail in the real clouds studied here. The \citet{draine09} model is
constrained by data from the very low density ISM ($A_V \lesssim$ a
few) whereas our sample of clouds is flux-limited to some very bright,
dense Galactic regions ($A_V > 20$). While modeling a ``molecular
cloud'', the \citet{bethell07} model bathes the cloud rather uniformly in a typical
interstellar radiation field which may be quite different from real
clouds containing embedded stars.

\subsection{Embedded Sources} \label{sec-embed}

\begin{figure*}
  \centering
  \plotone{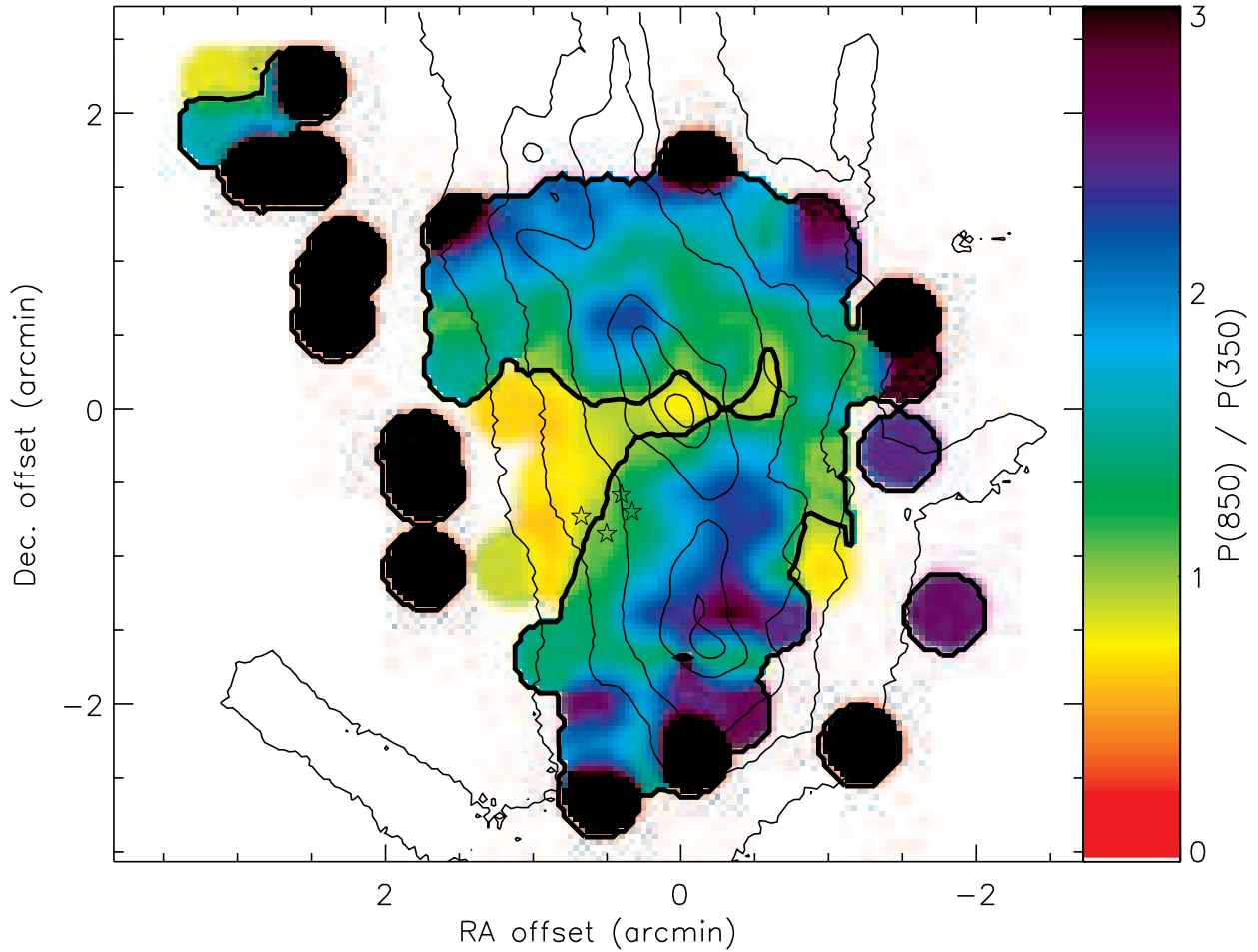}
  \caption{Map of the polarization ratio, $P(850)/P(350)$, in
    OMC-1. This map includes only $P\ge 3\sigma_p$ data but includes
    points typically rejected by the $\vert\Delta\phi\vert$ criterion
    discussed in the text.  Data with $P(850)/P(350)>3$ are shown as
    saturated (black) pixels.  Thin contours denote the $350\,\micron$
    intensity at levels of 1, 4, 8, 20, 40, and 80 \% of the peak intensity
    (data from SHARC-2; \citealt{omc1sharp}).  The thick contour is
    drawn at $P(850)/P(350)=1$.  For reference we also show the four
    Trapezium stars of M42. [\emph{A color version of this figure is
        available in the electronic version.}]}
  \label{fig-omc1pratio}
\end{figure*}

\begin{figure}
  \centering
  \includegraphics[height=0.7\textheight]{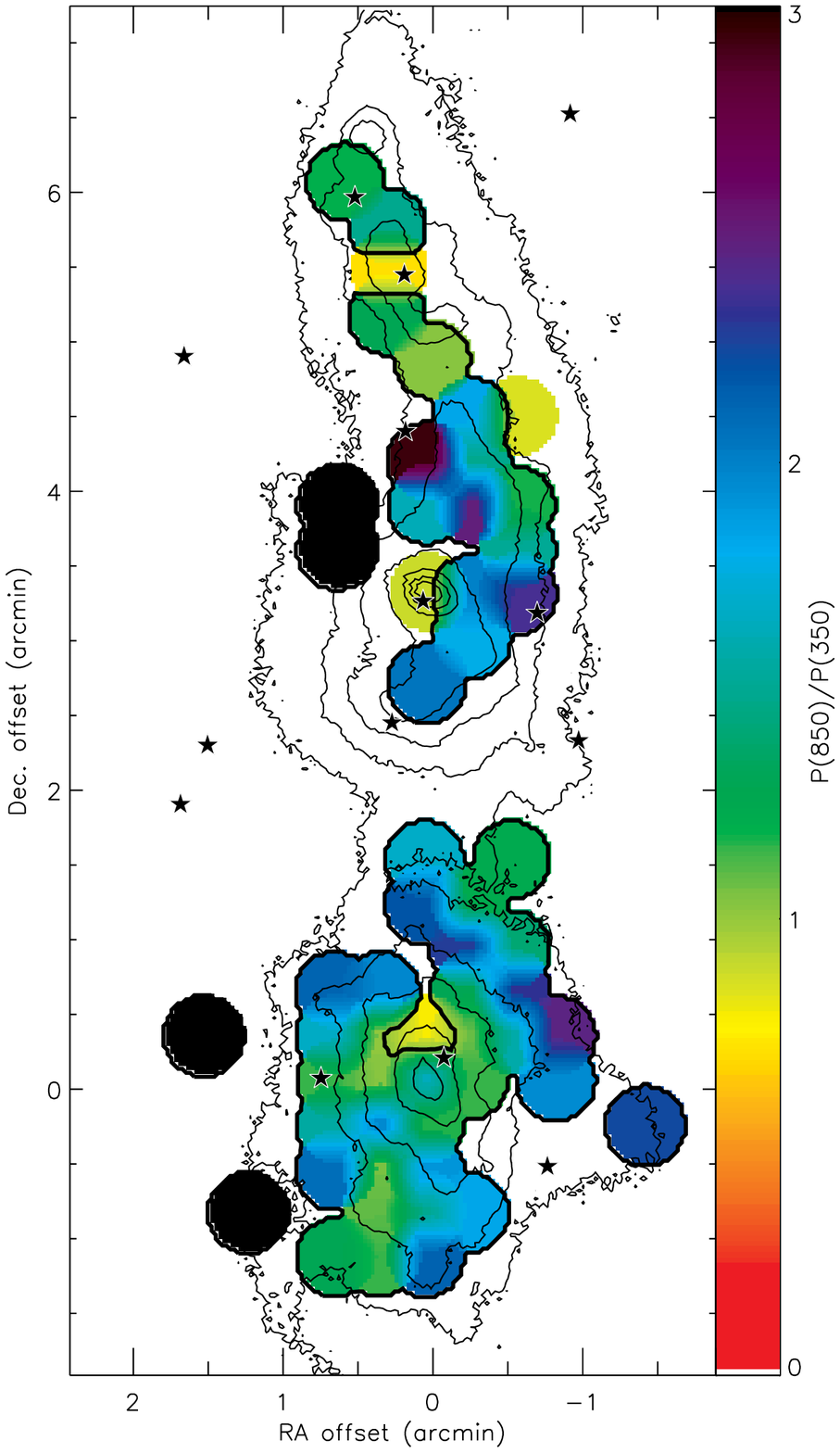}
  \caption{Same as Figure~\ref{fig-omc1pratio} but for DR21. Intensity
    data are from SHARC-2 (D. Dowell, private communication) Black
    stars indicate point sources identified by \emph{MSX}
    (Section~\ref{sec-embed}). [\emph{A color version of this figure
      is available in the electronic version.}]}
  \label{fig-dr21pratio}
\end{figure}

The RAT mechanism predicts that grains exposed to stronger radiation
sources will be more efficiently aligned.  From this we might expect
to see systematic trends in the polarization with distance from stellar
sources embedded in molecular clouds.  Such a trend is hinted at in
$60\,\micron$ polarization observations towards the W3A \ion{H}{2}
region \citep{w3}.  The most prominent embedded sources in OMC-1 are a
group of sources coincident with the BNKL intensity peak and the
Trapezium stars in the visible Orion nebula
(Fig.~\ref{fig-omc1pratio}). Using the MSX point-source
catalog\footnote{\url{http://irsa.ipac.caltech.edu/Missions/msx.html}}
\citep{price2001,msxpscv23} we have also identified a number of
embedded sources in the DR21 cloud (Fig.~\ref{fig-dr21pratio}).
However, the proximity of the sources to each other, coupled with the
fairly low spatial resolution of the polarization maps, does not allow
a careful quantitative study of the strength of the polarization (at
either wavelength) or the polarization ratio as a function of distance
from these sources.

\begin{figure}
  \centering
  \includegraphics[width=0.5\textwidth]{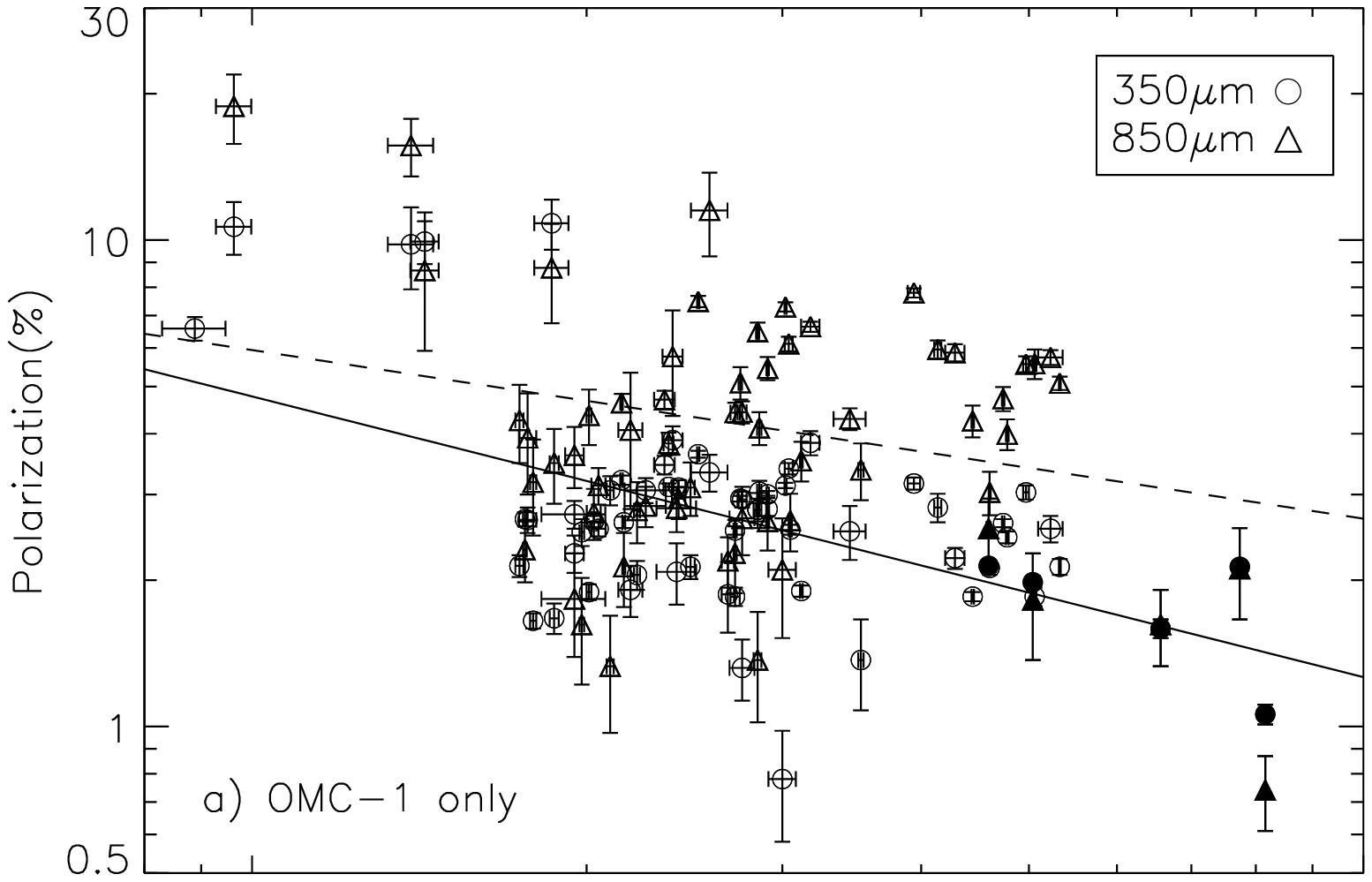}
  \includegraphics[width=0.5\textwidth]{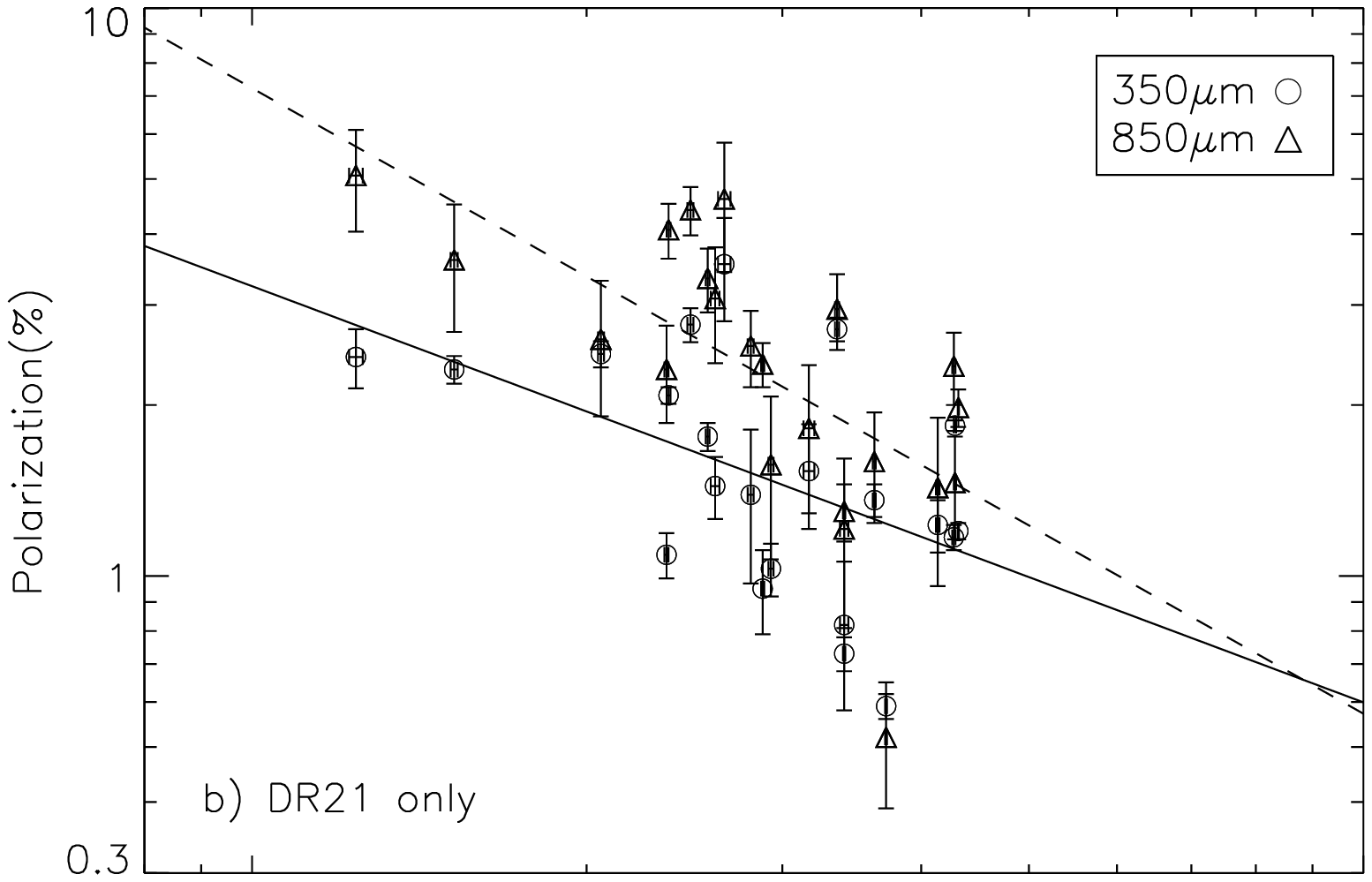}
  \includegraphics[width=0.5\textwidth]{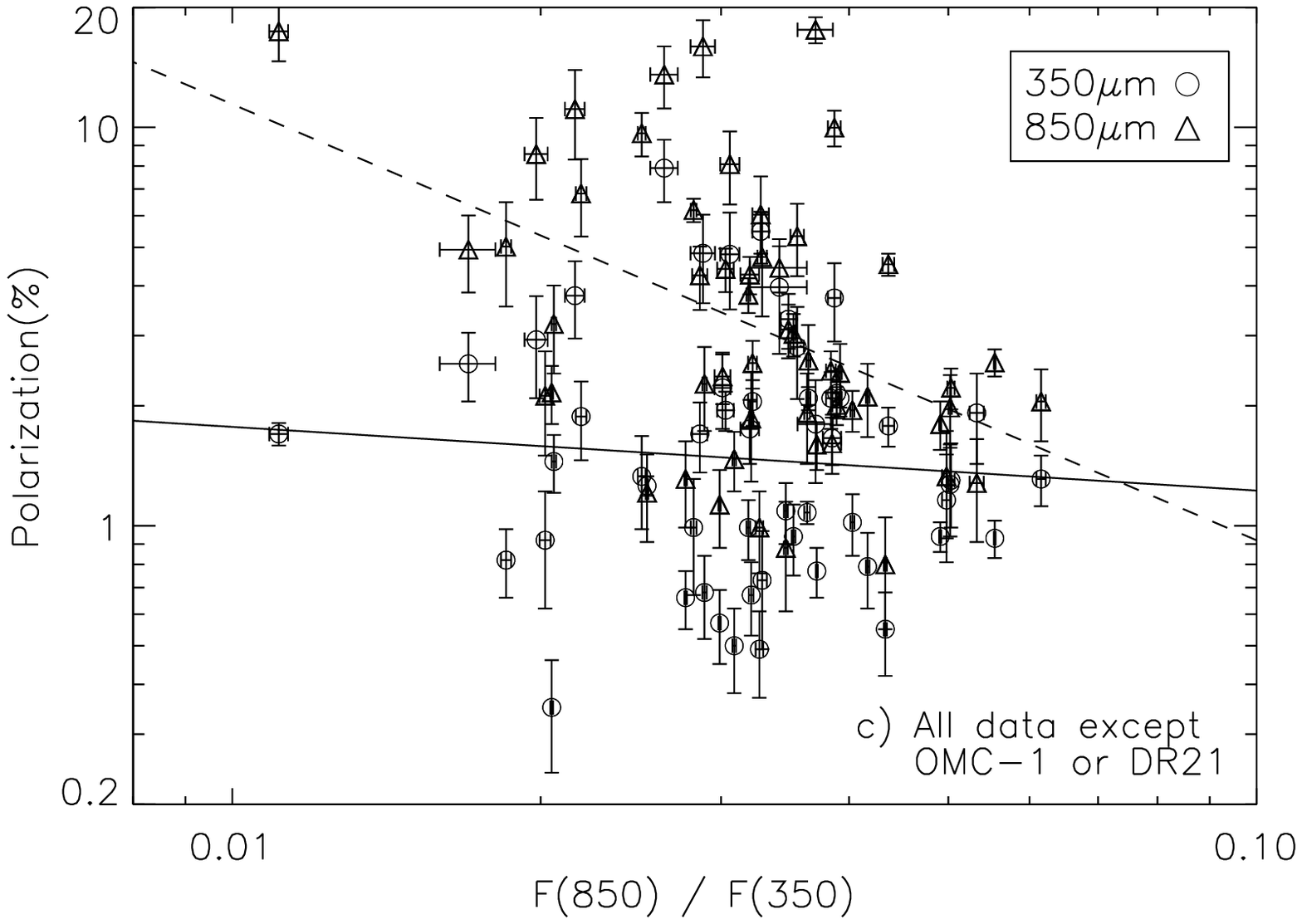}
  \caption{Polarization vs.\ intensity ratio in OMC-1 (\emph{a}), DR21
    (\emph{b}), and all data except OMC-1 and DR21 (\emph{c}).  Here
    we plot only data meeting the $P/\sigma_p$ and
    $\vert\Delta\phi\vert$ criteria described in the text.  The lines
    are power-law fits to the data (solid line for $350\,\micron$,
    dashed line for $850\,\micron$); fits do not consider error bars.
    In (\emph{a}) the fits exclude those points within $20\arcsec$ of
    OMC-1's central intensity peak (solid circles and solid triangles;
    see Section~\ref{sec-ratios}).  }
  \label{fig-colort}
\end{figure}

A trend in polarization efficiency with distance from a radiation
source also implies a correlation between the observed polarization
and dust temperature. A careful measure of the dust temperature
requires SED measurements over a wide range of wavelengths, a task
which is beyond the scope of the present work (e.g.,
\citealt{mythesis}).
%
To some extent, one can consider the
intensity or flux density ratio, $F(850)/F(350)$, as a proxy for the
temperature. 
Figures~\ref{fig-colort}\emph{a} and \ref{fig-colort}\emph{b} compare
the intensity ratio in OMC-1 and DR21 to the polarization at both 350
and 850 $\micron$. The polarization in both clouds generally drops
with increasing intensity ratio.  If we interpret these ratios as
color temperatures then the polarization increases with increasing
temperature, as would be expected for grains aligned via RAT\@.
However, we caution against over-interpretation of this result as
defining a color-temperature is problematic in the dense clouds for at
least two reasons.  First, the range plotted in
Figure~\ref{fig-colort} corresponds to unrealistically large
temperatures; assuming $\beta=2$ then $T=13$\,K for the largest ratio
$F(850)/F(350)=0.1$ is reasonable but $T>100$\,K for ratios
$F(850)/F(350)<0.033$. Secondly, the ratio may also be the result of
changes in grain emissivity (i.e., spectral index) and column density
as well as the temperature.

Using the intensity ratio as a proxy for temperature also has the
advantage that it is independent of distance, allowing us to combine
the relatively sparse data in individual clouds into a larger
dataset. Combining the remaining data in clouds other than DR21 or
OMC-1 results in Figure~\ref{fig-colort}\emph{c}. The same trend of
falling polarization is seen at both wavelengths. We emphasize that we
are comparing the polarization with the intensity \emph{ratio} and are
not discussing the ``polarization-hole" effect which is often observed
when comparing the polarization to absolute intensity at any given
wavelength (e.g., \citealt*{dasth,mfm02}).




%

\subsection{Intensity and Polarization Ratios} \label{sec-ratios}

One difficulty in using the absolute polarization values in
Section~\ref{sec-embed} above is that the observed polarization
magnitude is also a function of the parameters like the magnetic
field's LOS inclination angle, grain cross-section, and turbulence, all
of which may vary spatially across the cloud.
The inclination angle effect is mitigated somewhat by our choice to
limit the data set to those points with $\vert\Delta\phi\vert \le
10\arcdeg$ (see Section~\ref{sec-pratio}). These effects can be
further mitigated by using the ratio $P(850)/P(350)$. If the same
grains are responsible for the polarized emission at both wavelengths
then those ``polarization reduction'' factors effectively cancel in
the polarization ratio \citep{pspec}.

Figures~\ref{fig-omc1pratio} and \ref{fig-dr21pratio} show the spatial
distributions of the polarization ratio in OMC-1 and DR21,
respectively. Most of the mapped areas are characterized by
polarizations which are larger at $850\,\micron$ than $350\,\micron$.
Notable exceptions are intensity peaks in both objects (BNKL at the origin
of the OMC-1 map and DR21-OH(Main) at [$+3\farcs3,0$] in the DR21
map). The OMC-1 peak has also shown differences from the rest of the
cloud in other polarization work (e.g.,
\citealt{rao98,mythesis,omc1sharp}) so we will omit data within
$20\arcsec$ (one beam) of the peak in the analysis below.


%

\begin{figure}
  \centering
  \includegraphics[width=0.5\textwidth]{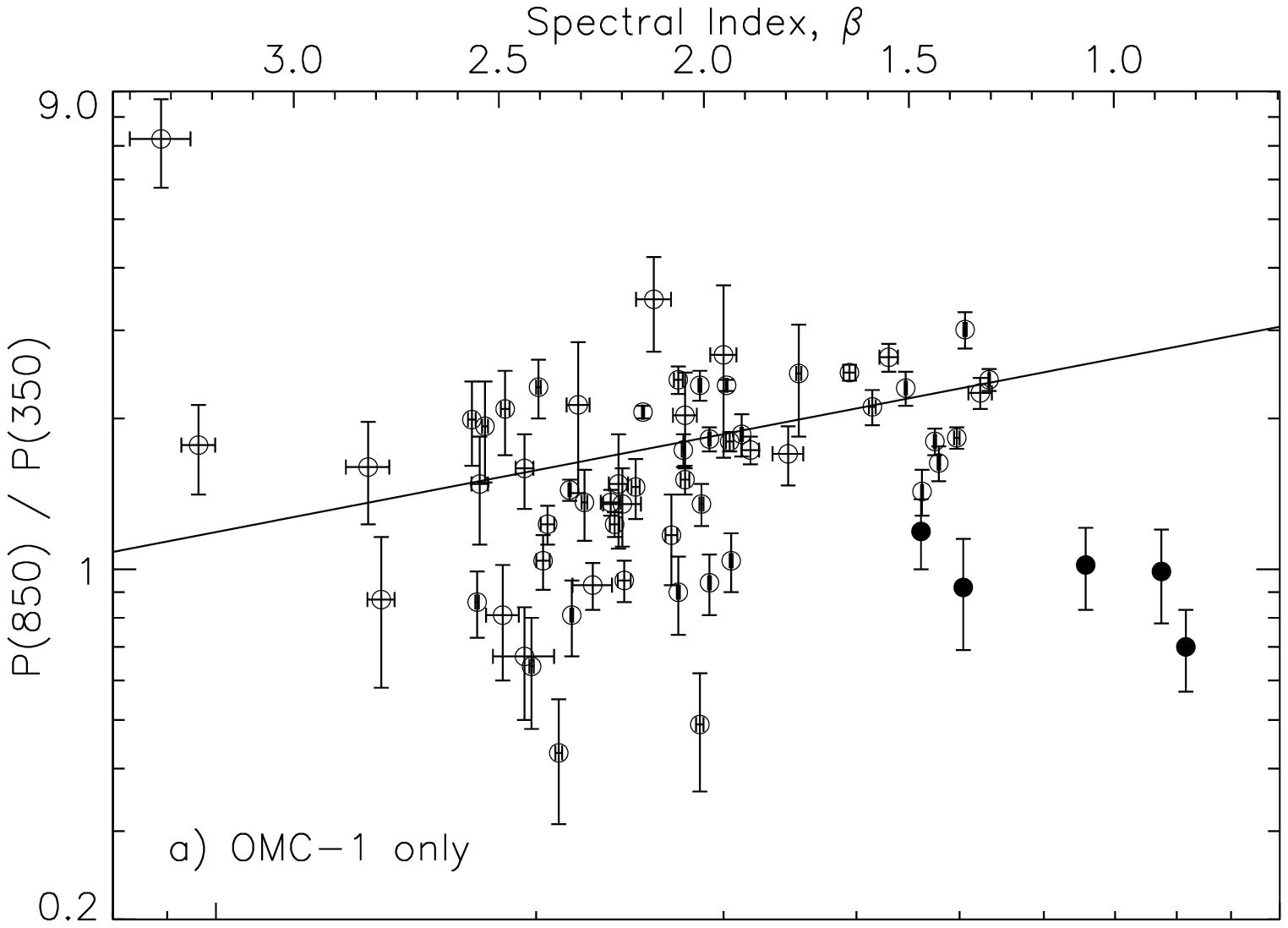}
  \includegraphics[width=0.5\textwidth]{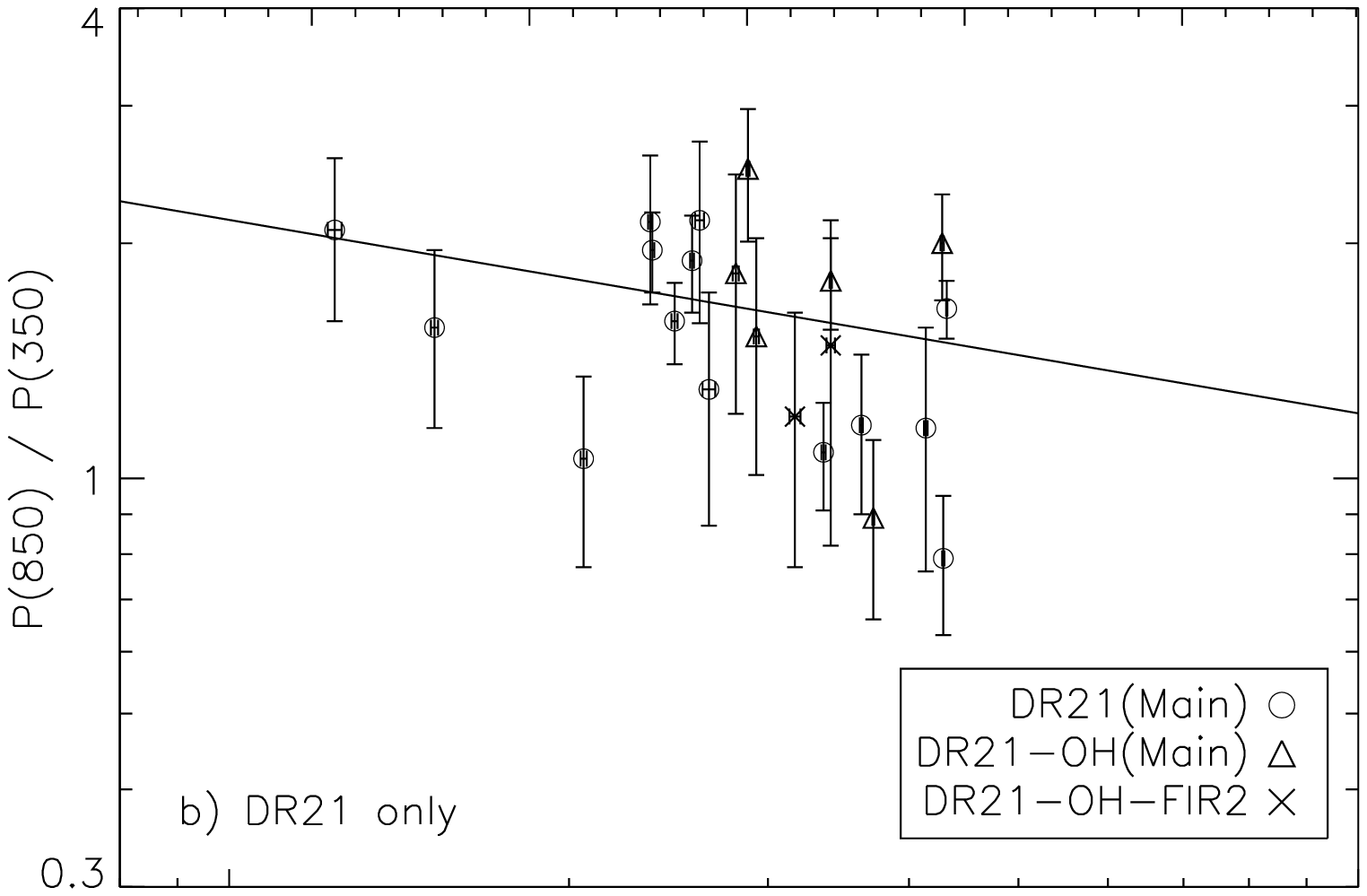}
  \includegraphics[width=0.5\textwidth]{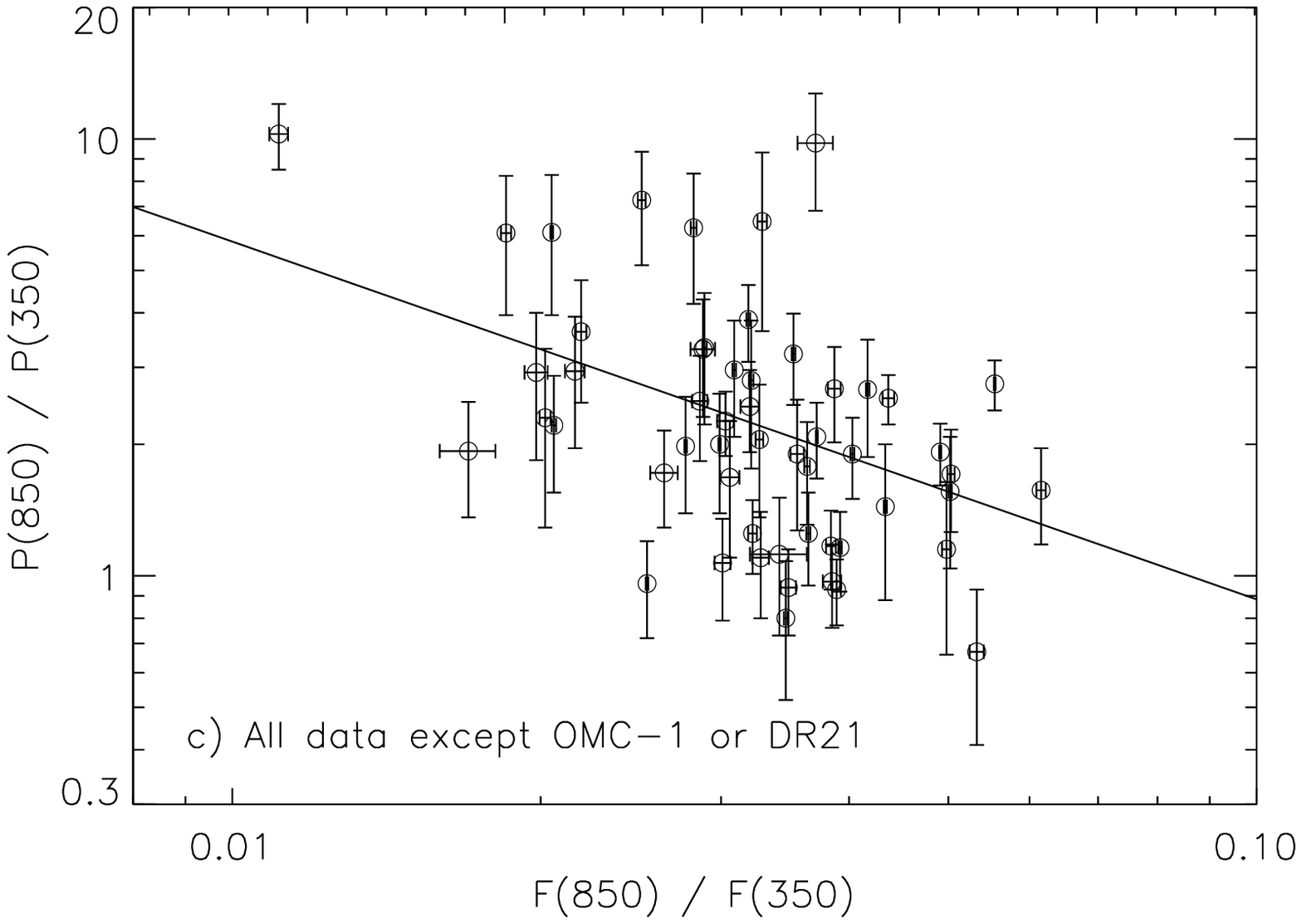}
  \caption{Polarization and intensity ratios in OMC-1 (\emph{a}), DR21
    (\emph{b}), and all data except OMC-1 and DR21 (\emph{c}). The
    upper $x$-axis scales cast the intensity-ratio as the standard spectral
    index defined through the relation
    $F(\nu)\propto\nu^{\beta+2}$. Here we plot only data meeting the
    $P/\sigma_p$ and $\vert\Delta\phi\vert$ criteria described in the text.  The lines are
    power-law fits to the data;
    in (\emph{a}) the fits exclude those points within $20\arcsec$ of
    OMC-1's central intensity peak (solid circles; see
    Section~\ref{sec-ratios}).  }
  \label{fig-pvsbeta}
\end{figure}

The most direct tests of the grain alignment models using
submillimeter data require comparisons between the measured
polarization ratio and the dust temperatures, spectral indices, and/or
radiation environment of the aligned grains.  A careful measure of
those parameters requires SEDs measured over a wide range of
wavelengths, a task which is beyond the scope of the present work
(e.g., \citealt{mythesis}).  However, we can again use the intensity
ratio, $F(850)/F(350)$, as a proxy for the temperature or spectral
index.  Very different trends are observed when comparing this ratio
to the polarization ratio in OMC-1 and DR21
(Figures~\ref{fig-pvsbeta}\emph{a} and \ref{fig-pvsbeta}\emph{b}).
The trend is an increase in $P(850)/P(350)$ in OMC-1 but a decrease in
DR21.  
As before the intensity ratio is independent of distance, allowing us
to combine the remaining data in clouds other than DR21 or OMC-1
(Fig.~\ref{fig-pvsbeta}\emph{c}).  No strong trend between the
intensity ratio and polarization ratio is observed.

The large amount of scatter in these observations is not unexpected.
The dense clouds studied here are certainly composed of multiple
temperature components covering a wide range (e.g., $\sim 20$ -- 80
K\@). Therefore, the intensity ratio is a measure not only of the
components' dust temperatures, but also their different spectral
indices and relative column densities. Presumably this effect is
partly the cause of the large scatter observed in the data of
Figure~\ref{fig-pvsbeta}.

\section{Summary}


We have compiled all the spatially coincident data available from the
Hertz ($350\,\micron$; \citealt{hertzarchive}) and SCUBA
($850\,\micron$; \citealt{scubaarchive}) polarimeters. In order to
facilitate this comparison the SCUBA-pol data have been spatially
re-sampled to match the spatial locations of the Hertz data; the two
instruments have comparable spatial resolution ($20\arcsec$).  We find
a total of 1699 individual locations which can be compared within 17
different Galactic clouds; all the polarization and intensity data at
350 and $850~\micron$ are given in electronic format in Table~2. Of
these data 1124 points yield non-zero polarizations at both
wavelengths when corrected for the positive polarization bias.
Additionally, of the bias-corrected data, 398 points satisfy the
criterion $P/\sigma_p \ge 3$ at both wavelengths, and 141 points
satisfy the additional criterion $|\phi(850)-\phi(350)| \le
10\arcdeg$. Complete polarization and intensity maps for all clouds
are given in the Appendix.

We have investigated the change in polarization angle from
850-to-$350~\micron$.  The angle differences exhibit a wide
distribution indicating that, in some regions, there is a real angle
rotation within the measurement uncertainties.  However, the angle
distributions are centered about $\phi(350)\approx\phi(850)$.  These
conclusions hold globally for the entire $P/\sigma_p \ge 3$ data set
in this work and individually for the $P/\sigma_p \ge 3$ data sets in
the clouds OMC-1, OMC-3, and DR21. Due to the limited number of data
points in other individual clouds we have made no attempt to study the
angle distributions nor the point-by-point angle agreement between
wavelengths; therefore these conclusions do not necessarily extend to
the other clouds included herein.

We have also examined the 850-to-$350~\micron$ polarization ratio,
$P(850)/P(350)$, on a point-by-point basis at every spatial location
in each of our sampled clouds.  From this work we establish a genuine
trend towards higher polarization at $850\,\micron$ than
$350\,\micron$ with a median polarization ratio of $P(850)/P(350) =
1.7$ (and a median absolute deviation of 0.6).  These values cover all
clouds in our dataset and points which satisfy the criteria
$P/\sigma_p \ge 3$ and also $|\phi(850)-\phi(350)| \le 10\arcdeg$.
This trend is consistent with previous work (e.g.,
\citealt{pspec,mythesis,omc1sharp}) and is best explained by models
which require mixtures of dust grains with different physical
properties (i.e., temperatures and spectral indices) and different
alignment efficiencies.  The dust and alignment models of
\citet{bethell07} and \citet{draine09} predict increasing values for
the polarization from 350-to-$850~\micron$ ($\sim1.0$--1.3) but not of
the same magnitude as observed here.  This is most likely due to the
fact that their models use very different physical conditions than
prevail in our sample of bright, dense molecular clouds.

We find a trend in which the 350 and 850 $\micron$ polarizations tend
to fall as the 850-to-350 $\micron$ intensity ratio increases.  If we
interpret this ratio as a color temperature then these data are
consistent with a key prediction of radiative alignment torques.  That
is, grains which are more exposed to radiation sources, and are thus
warmer, are more efficiently aligned.  However, we caution that the
assignment of temperature to a two-wavelength intensity ratio is not
robust. No clear trends are observed when the polarization ratios are
compared to the intensity ratios on a point-by-point basis.  Better
tests require work at additional wavelengths in order to produce SEDS
from which more accurate dust temperatures can be extracted.

\acknowledgements This work would not have been possible without the
dedicated support of the staff at the Caltech Submillimeter
Observatory and James Clerk Maxwell Telescope and the extended team of
dedicated scientists who toiled to collect and analyze data for both
Hertz and SCUBA-pol.  We would like to thank Roger Hildebrand, Giles
Novak, and B-G Andersson for comments on an early draft of this
paper. This research has made use of the NASA/\-IPAC Infrared Science
Archive, which is operated by the Jet Propulsion Laboratory,
California Institute of Technology, under contract with the National
Aeronautics and Space Administration.

\clearpage

\appendix

\section{Polarization Maps} \label{sec-maps}

\begin{figure}
  \includegraphics[width=0.9\columnwidth]{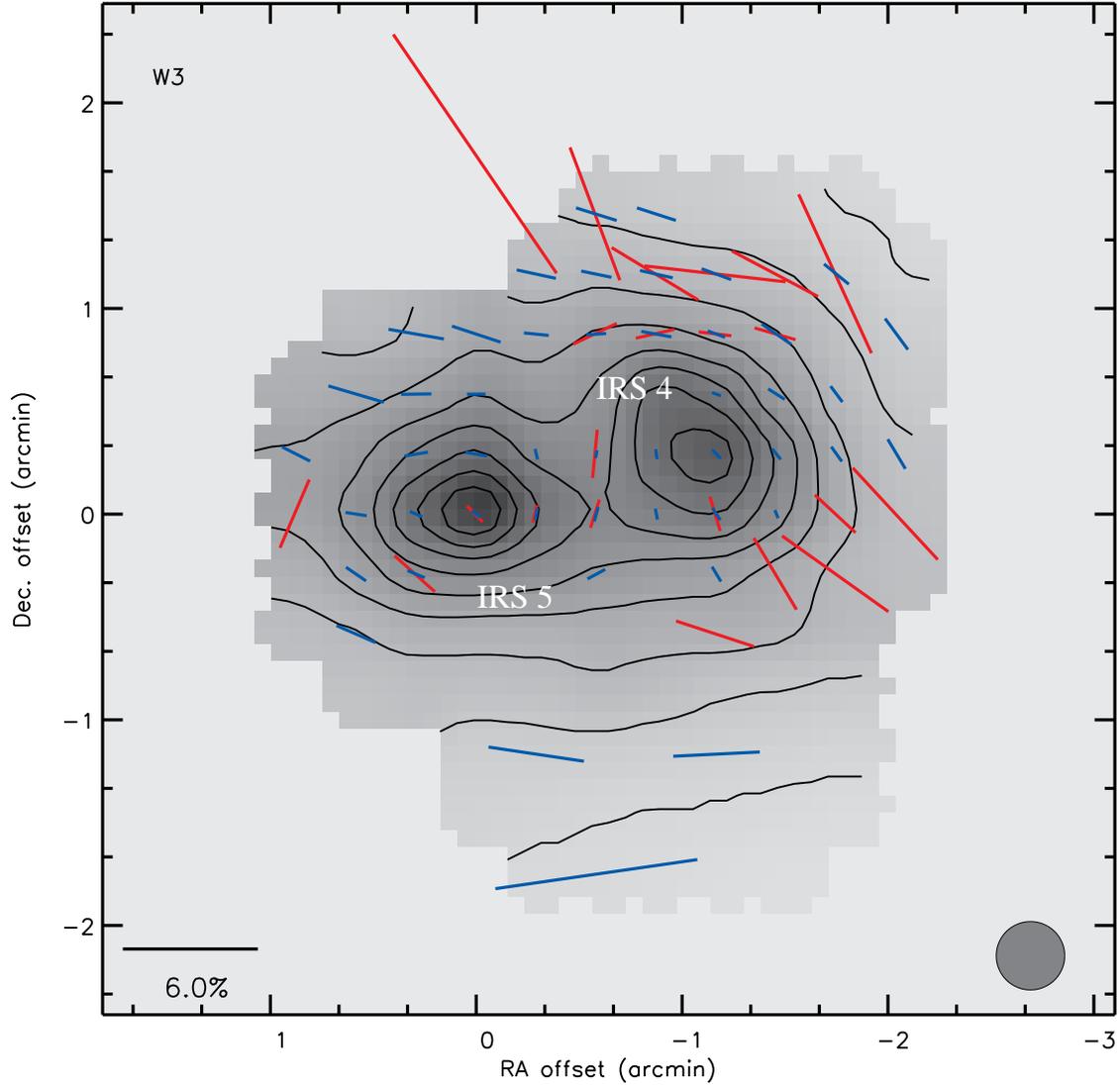}
  \caption{W3. The intensity map is from Hertz at $350\,\micron$ with
    contours drawn at 10, 20, 30, ..., 90\% of the peak
    intensity. Other key map features for this and subsequent figures
    are described in the text.}
  \label{fig-map1}
\end{figure}

\begin{figure}
  \includegraphics[width=0.9\columnwidth]{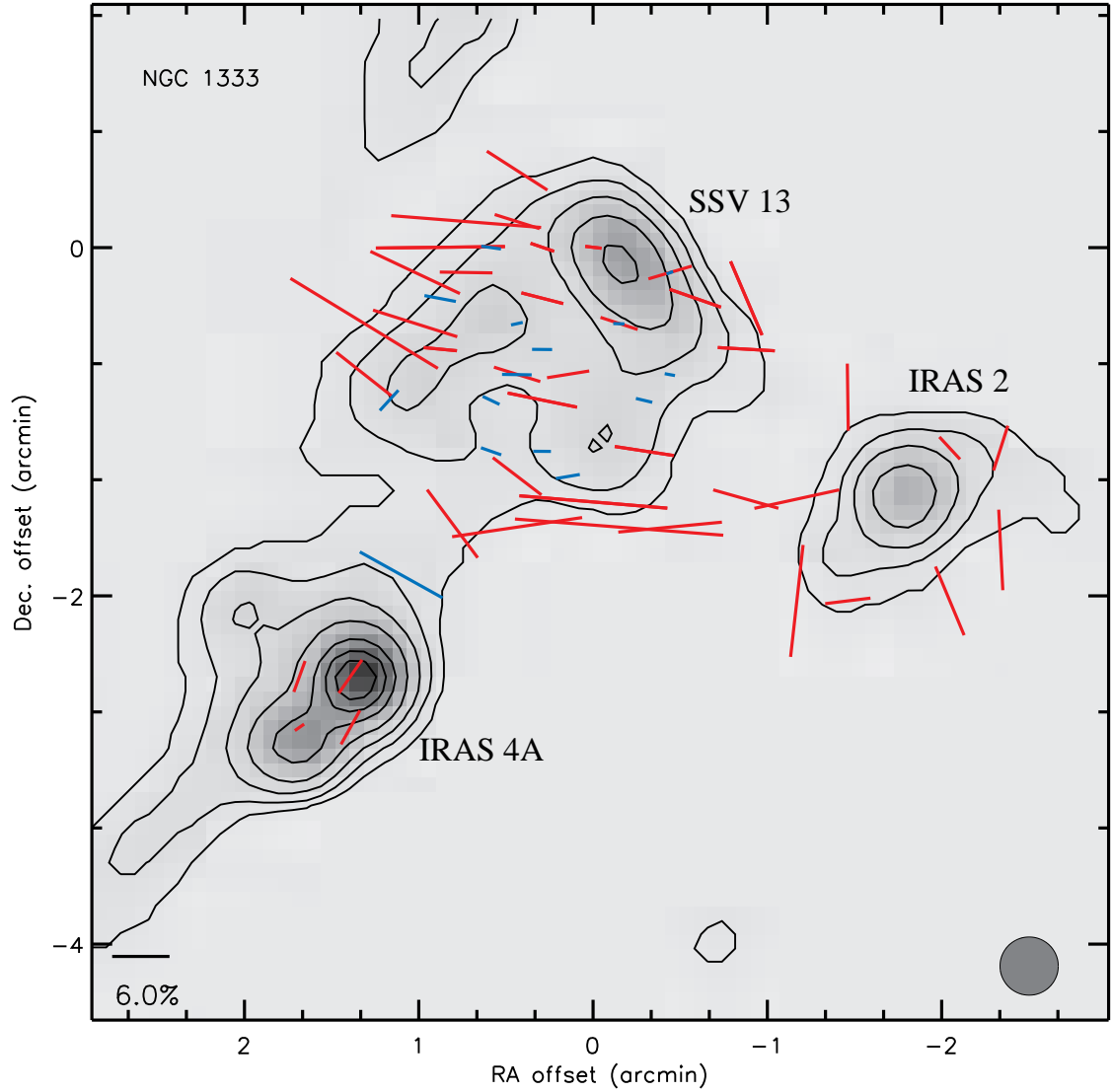}
  \caption{NGC\,1333.
    The intensity map is from SCUBA with contours drawn at 2, 5, 10,
    20, 40, 60, and 80\% of the peak $850\,\micron$ intensity (which
    occurs in IRAS\,4A).}
  \label{fig-map2}
\end{figure}

\begin{figure*}
  \includegraphics[width=0.9\textwidth]{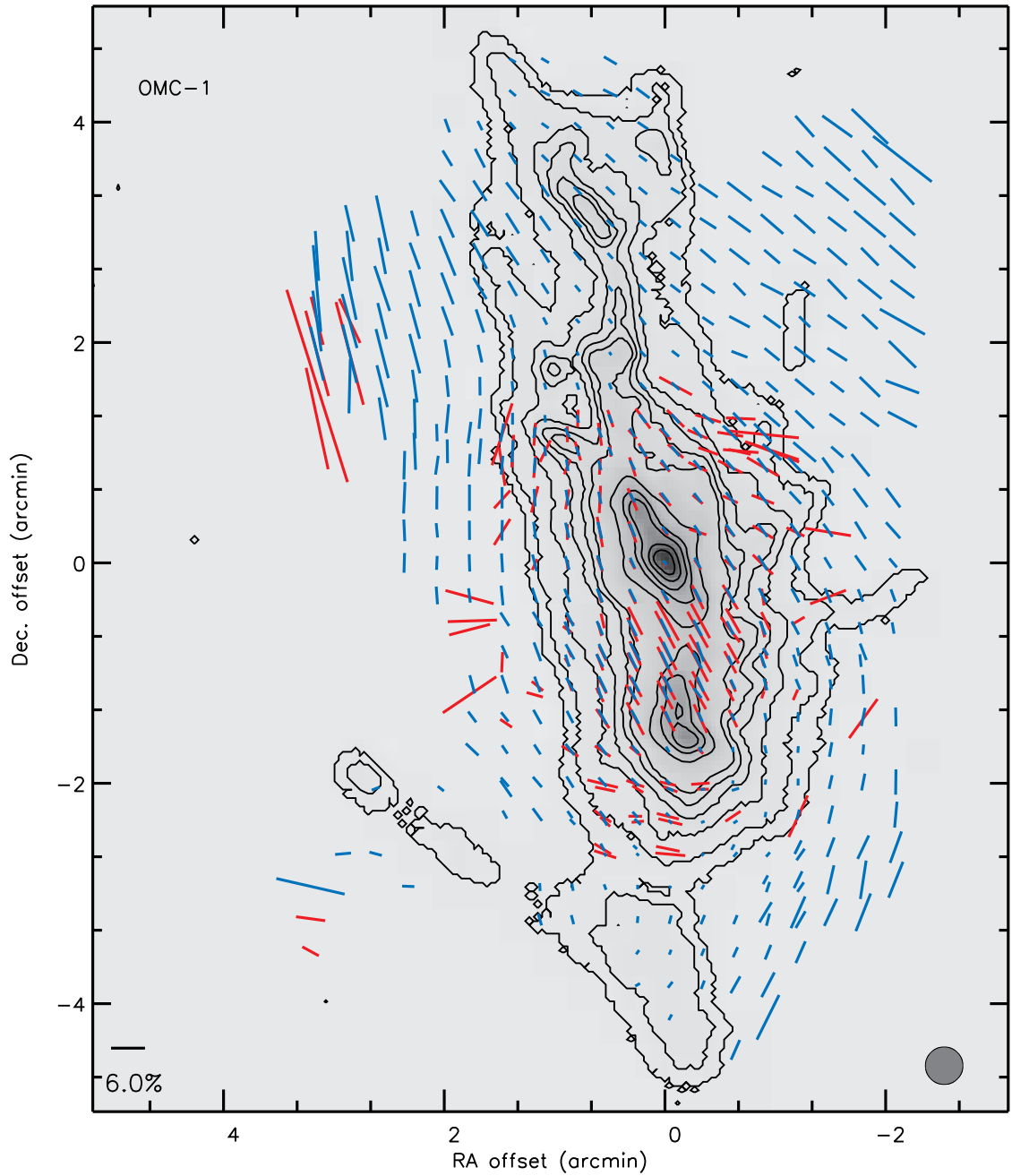}
  \caption{OMC-1. The intensity map is from SHARC-2 \citep{omc1sharp}
    with contours drawn at 1, 2, 4, 6, 8, 10, 20, 40, 60, and 80 \% of
    the peak $350\,\micron$ intensity. The gray circle in the
    lower-right indicates the $20\arcsec$ effective beam-size of Hertz
    and SCUBA-pol, not the SHARC-2 intensity data which has a beam
    size of $\sim$10\arcsec.}
  \label{fig-map3}
\end{figure*}

\begin{figure}
  \includegraphics[width=0.9\columnwidth]{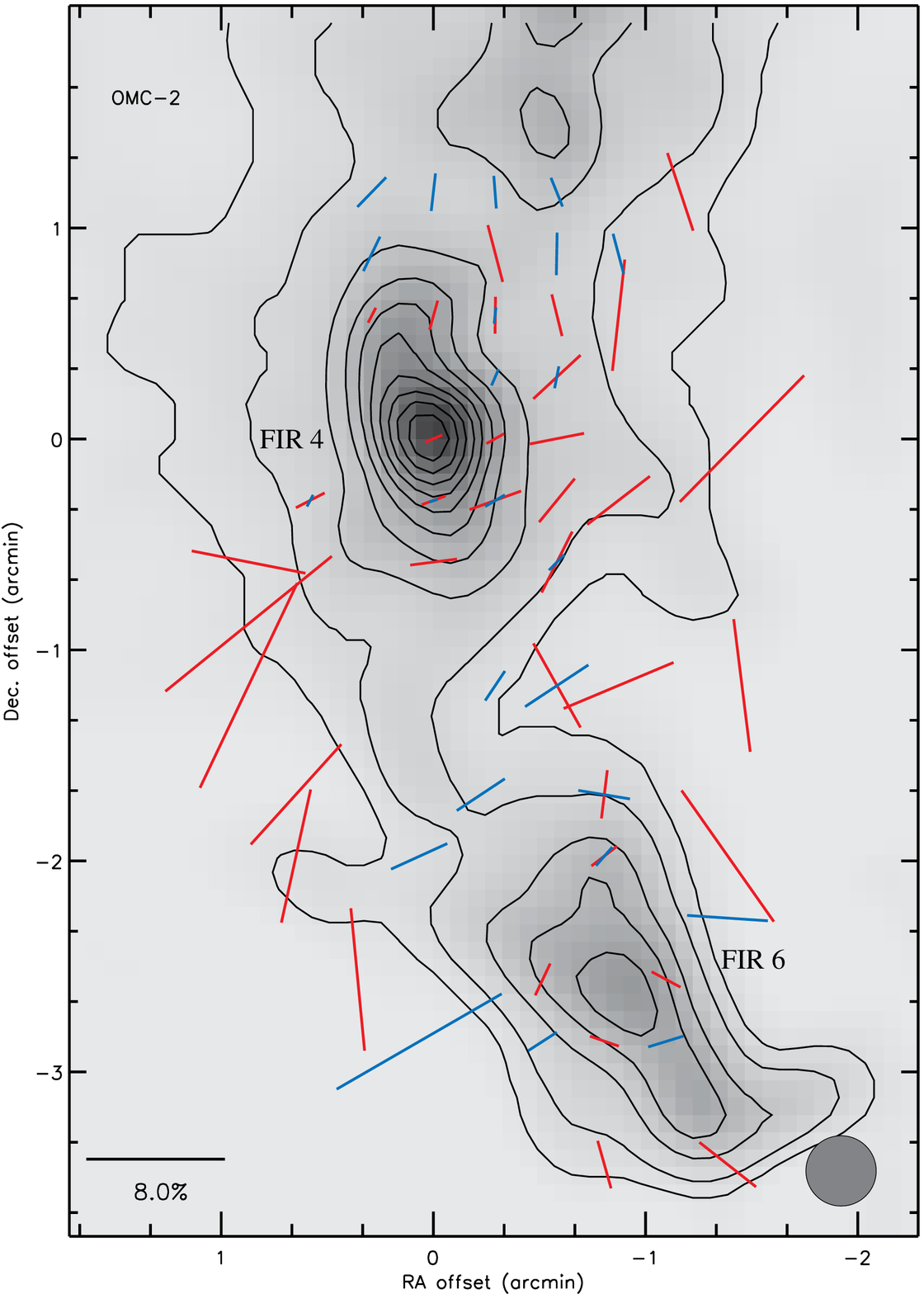}
  \caption{OMC-2. The intensity map is from SCUBA with contours drawn
    at 5, 10, 20, 30, ..., 90\% of the peak $850\,\micron$
    intensity. 
}
  \label{fig-map4}
\end{figure}

\begin{figure}
  \centering
  \includegraphics[width=0.7\columnwidth]{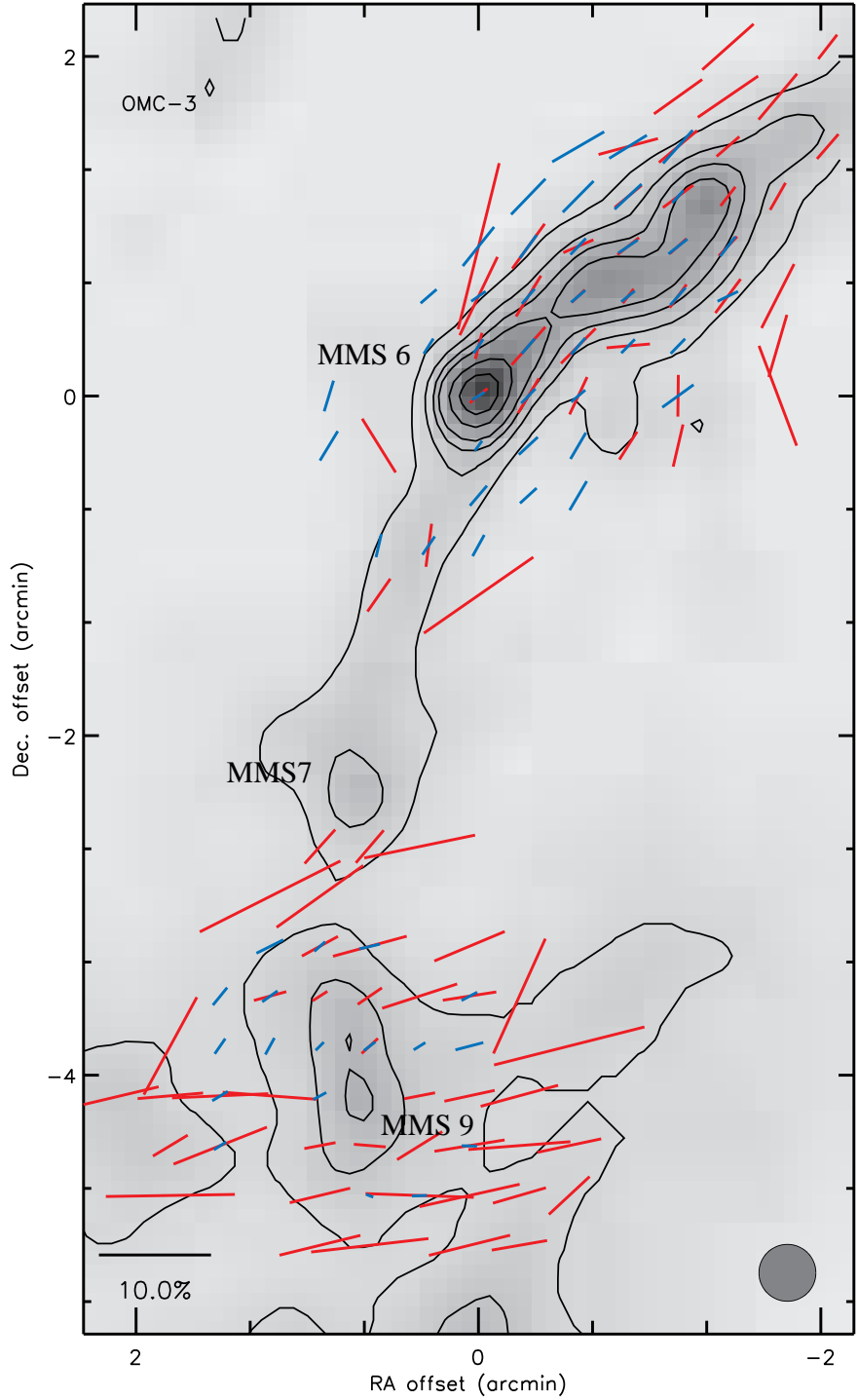}
  \caption{OMC-3.  The SCUBA intensity is shown with contours at 10,
    20, 30, 40, 60, and 80\% of the peak $850\,\micron$ intensity. 
    (Note that some SCUBA-pol vectors in the center of the map have been
    removed where there was no Hertz-coverage of this cloud.)}
  \label{fig-map5}
\end{figure}

\begin{figure}
  \includegraphics[width=0.9\columnwidth]{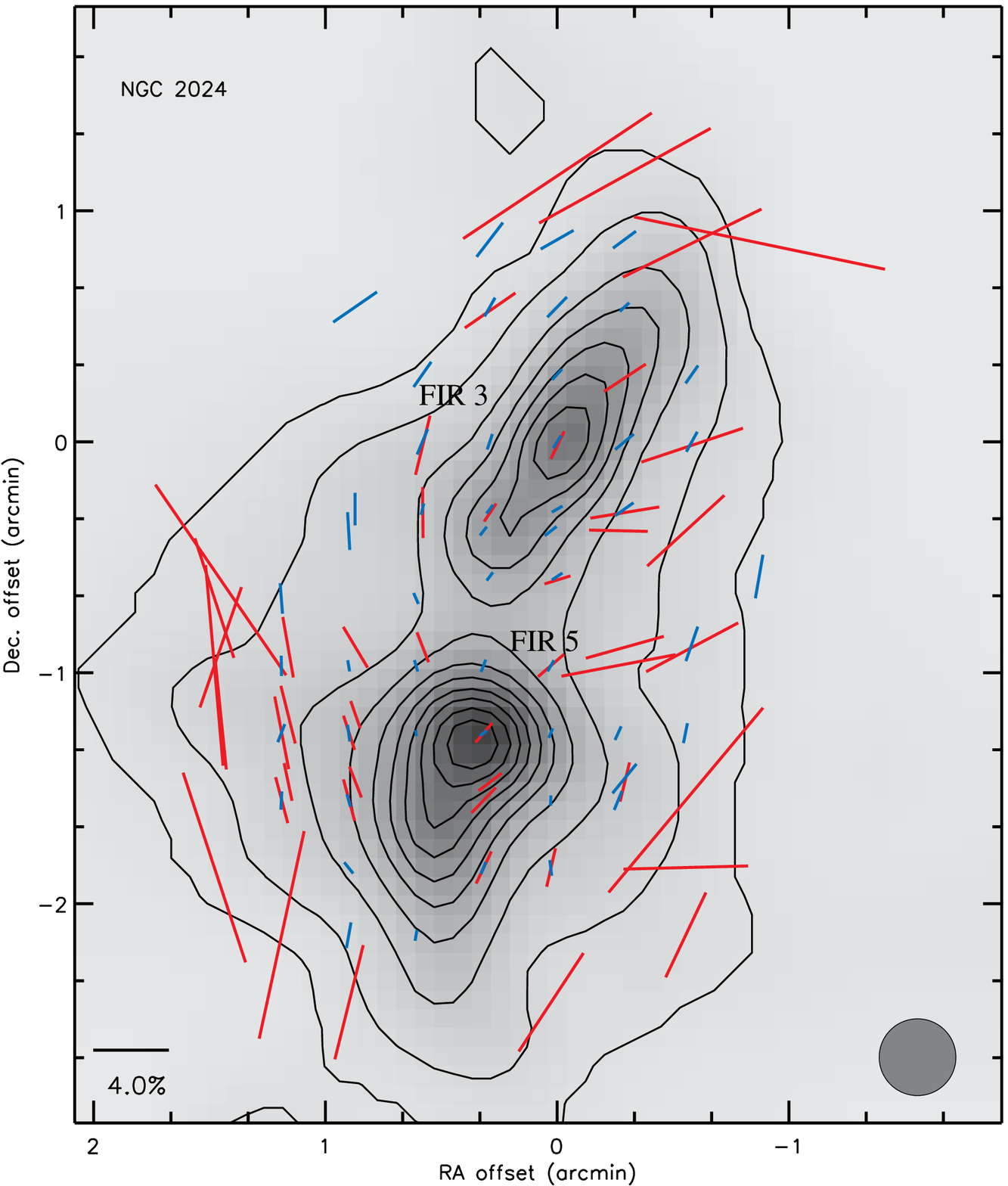}
  \caption{NGC\,2024. The intensity map is from SCUBA with contours
    drawn at 5, 10, 20, 30, ..., 90\% of the peak $850\,\micron$
    intensity. 
}
  \label{fig-map6}
\end{figure}

\begin{figure}
  \includegraphics[width=0.9\columnwidth]{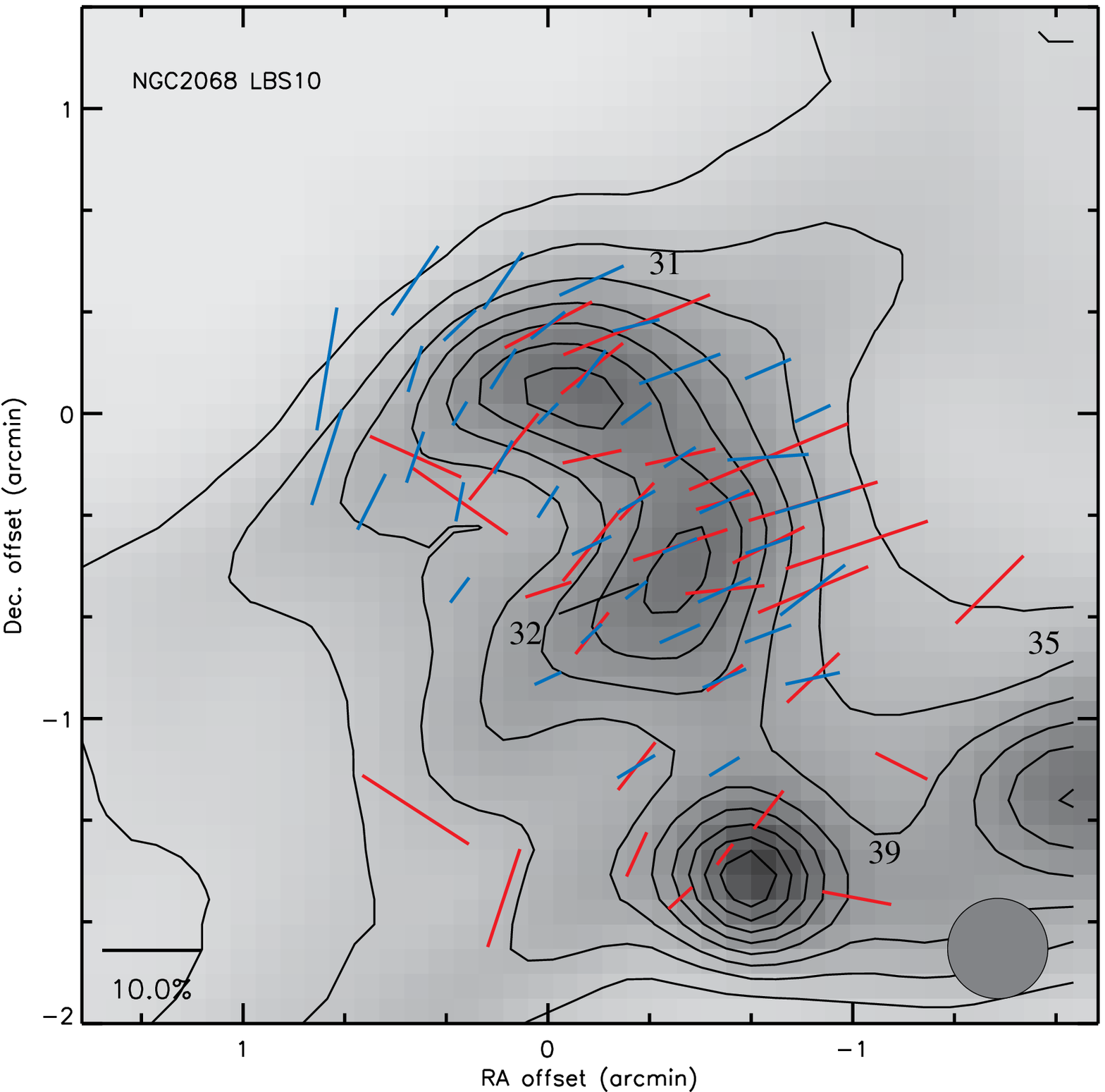}
  \caption{NGC\,2068 LBS\,10. The intensity map is from SCUBA with
    contours drawn at 10, 20, 30, ..., 90\% of the peak $850\,\micron$
    intensity. Cloud core labels are from \citet{mitchell2001}.}
  \label{fig-map7}
\end{figure}

\begin{figure}
  \includegraphics[width=0.9\columnwidth]{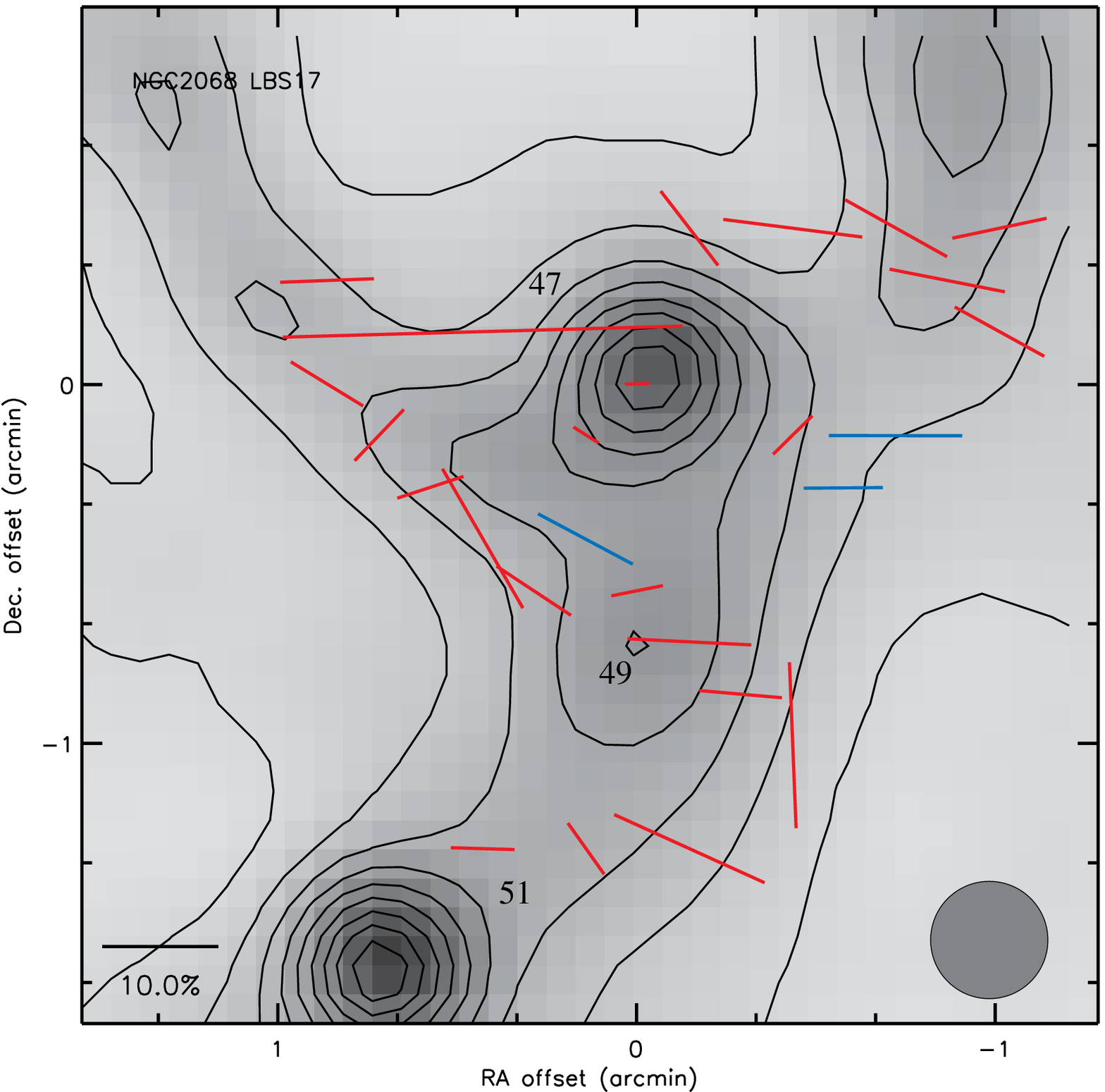}
  \caption{NGC\,2068 LBS\,17. The intensity map is from SCUBA with
    contours drawn at 10, 20, 30, ..., 90\% of the peak $850\,\micron$
    intensity.  Cloud core labels are from \citet{mitchell2001}.}
  \label{fig-map8}
\end{figure}

\begin{figure}
  \includegraphics[width=0.9\columnwidth]{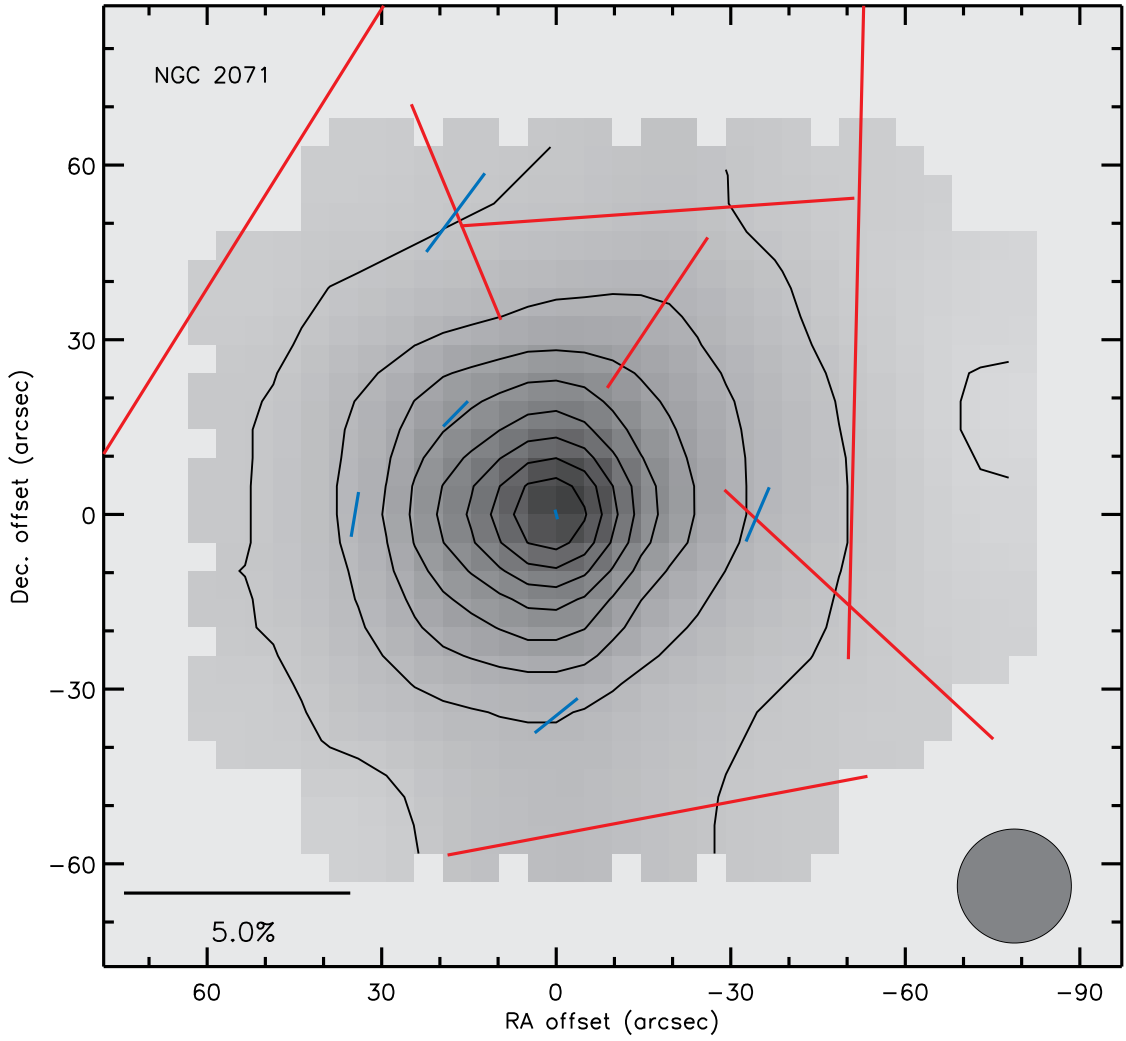}
  \caption{NGC\,2071. The intensity map is from Hertz with
    contours drawn at 20, 30, ..., 90\% of the peak $350\,\micron$
    intensity.}
  \label{fig-map8.5}
\end{figure}

\begin{figure}
  \includegraphics[width=0.9\columnwidth]{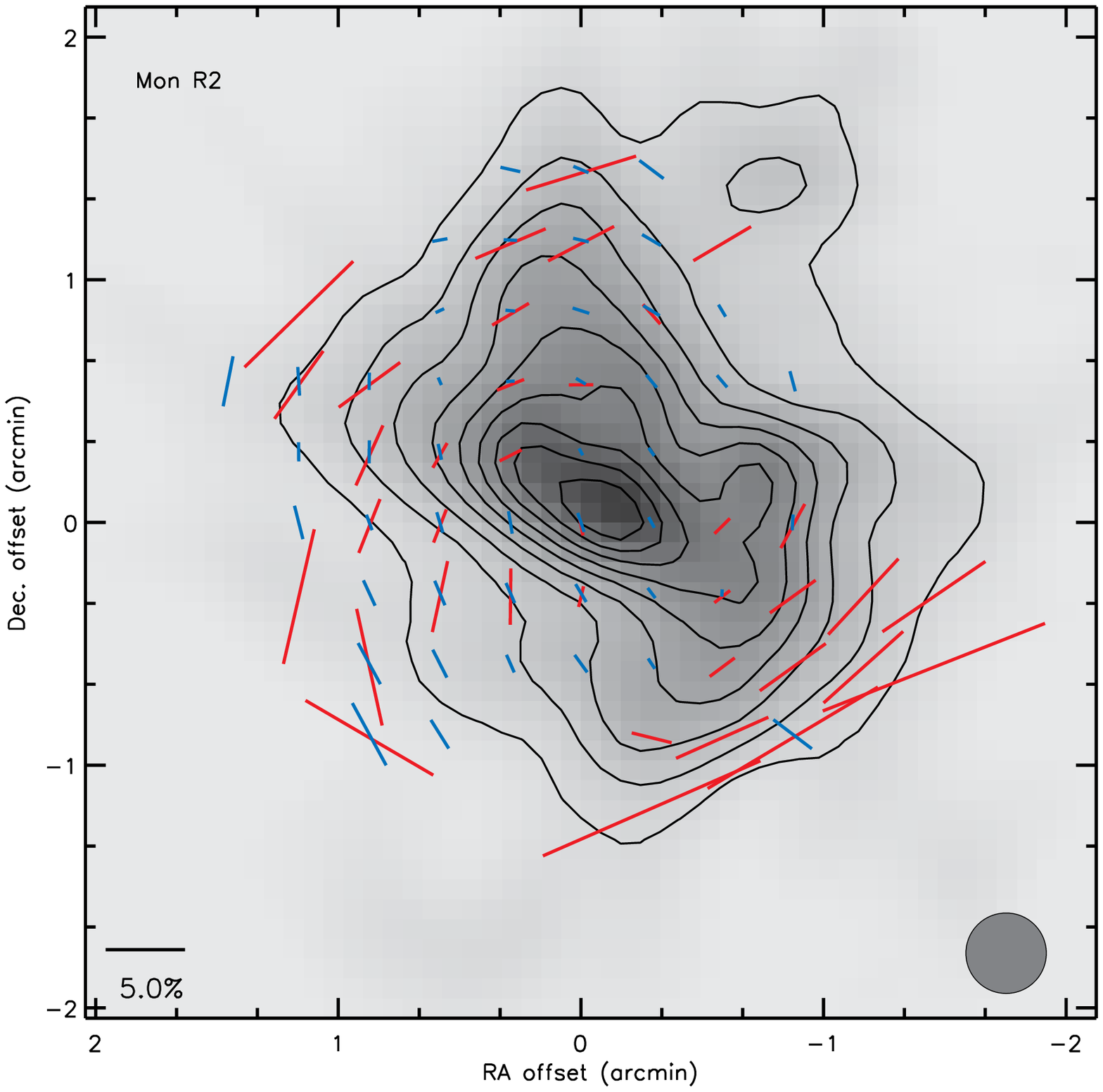}
  \caption{Mon\,R2. The intensity map is from SCUBA with
    contours drawn at 10, 20, 30, ..., 90\% of the peak $850\,\micron$
    intensity.}
  \label{fig-map9}
\end{figure}

\begin{figure}
  \includegraphics[width=0.9\columnwidth]{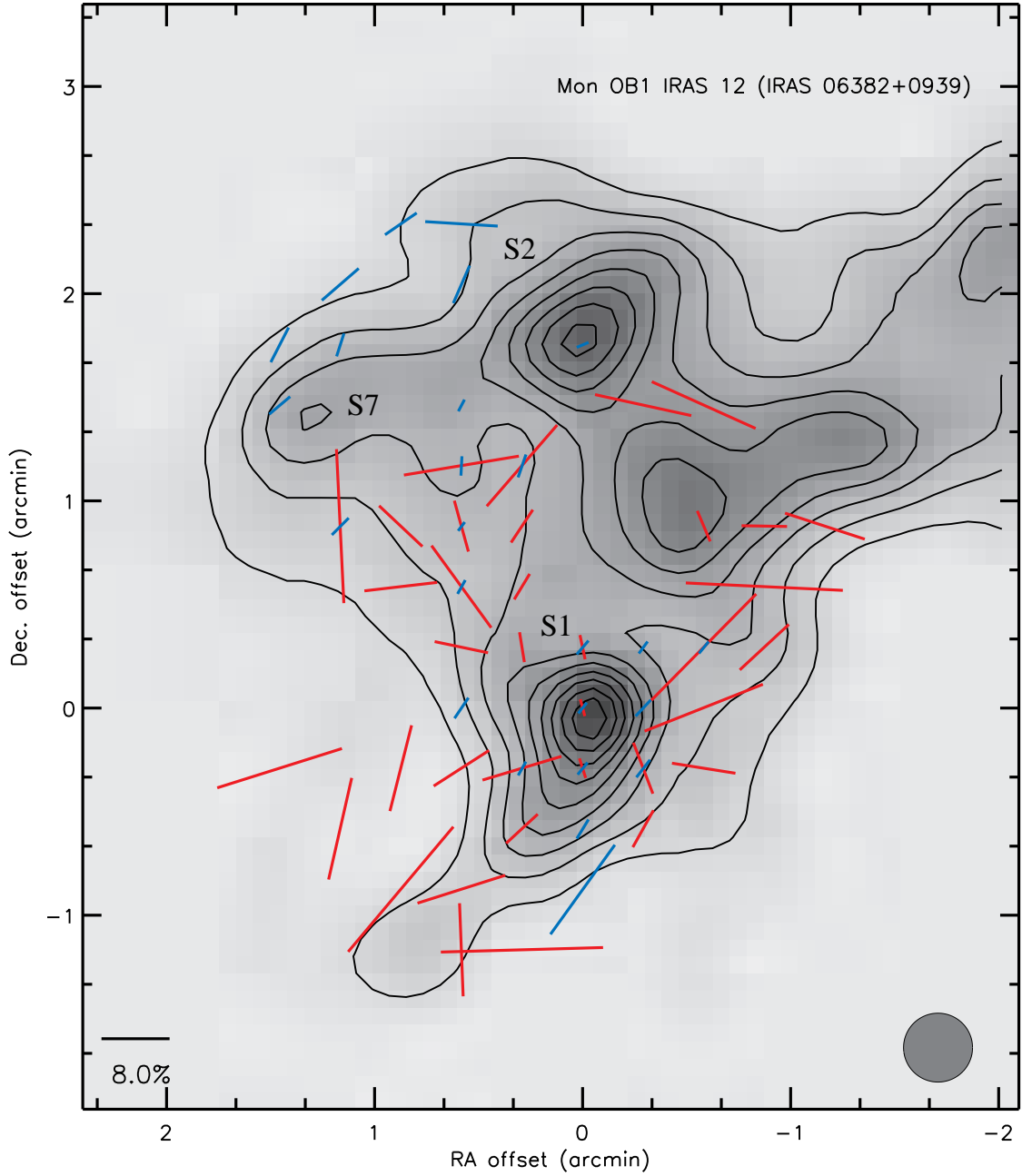}
  \caption{Mon\,OB1 IRAS\,12 (IRAS 06382+0939). The intensity map is
    from SCUBA with contours drawn at 10, 20, 30, ..., 90\% of the
    peak $850\,\micron$ intensity. Cloud core labels are from
    \citet{wolfchase2003}}
  \label{fig-map10}
\end{figure}

\begin{figure}
  \includegraphics[width=0.9\columnwidth]{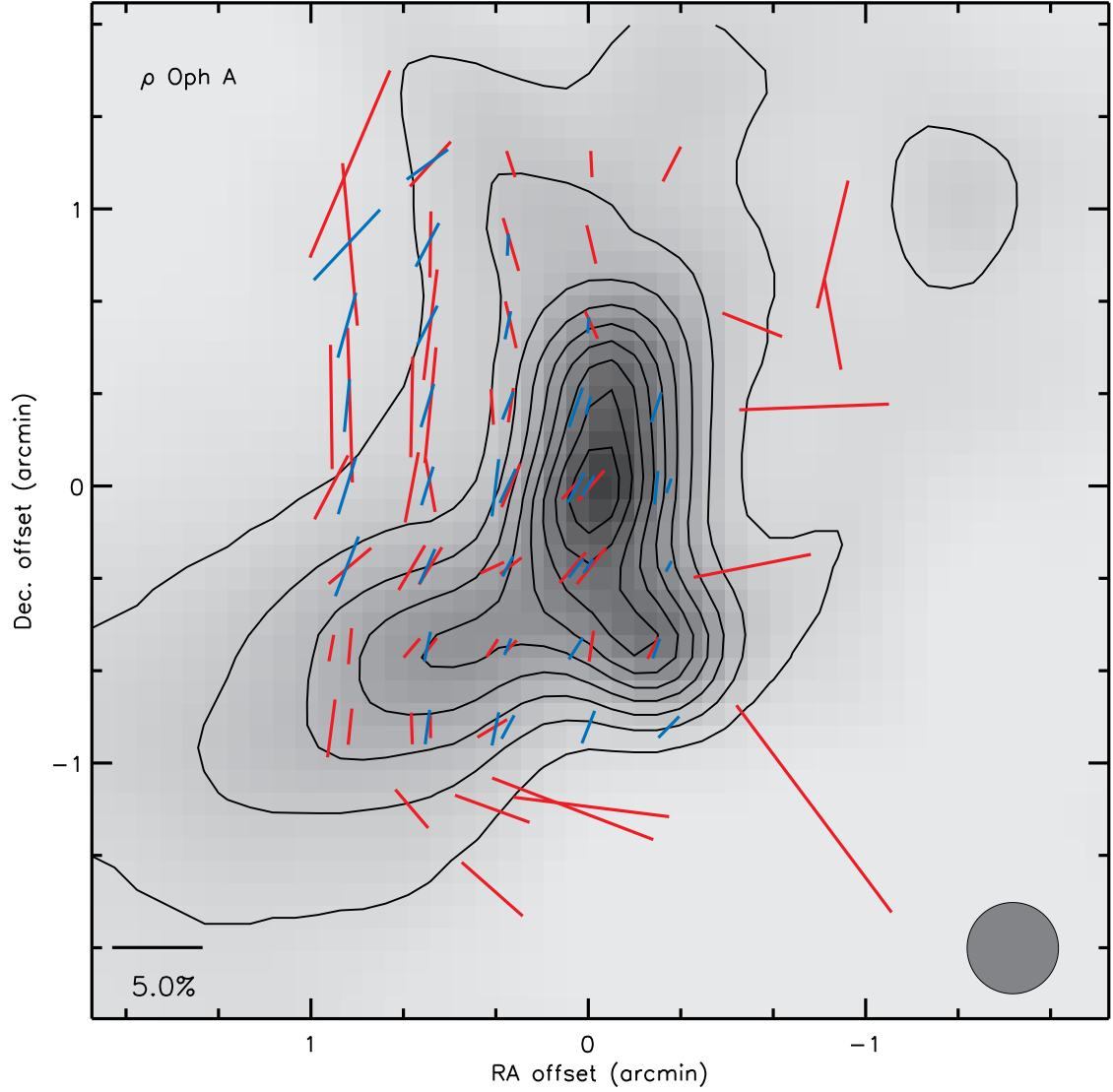}
  \caption{$\rho$\,Oph A.  The intensity map is from SCUBA with
    contours drawn at 10, 20, 30, ..., 90\% of the peak $850\,\micron$
    intensity.}
  \label{fig-map11}
\end{figure}

\begin{figure}
  \includegraphics[width=0.9\columnwidth]{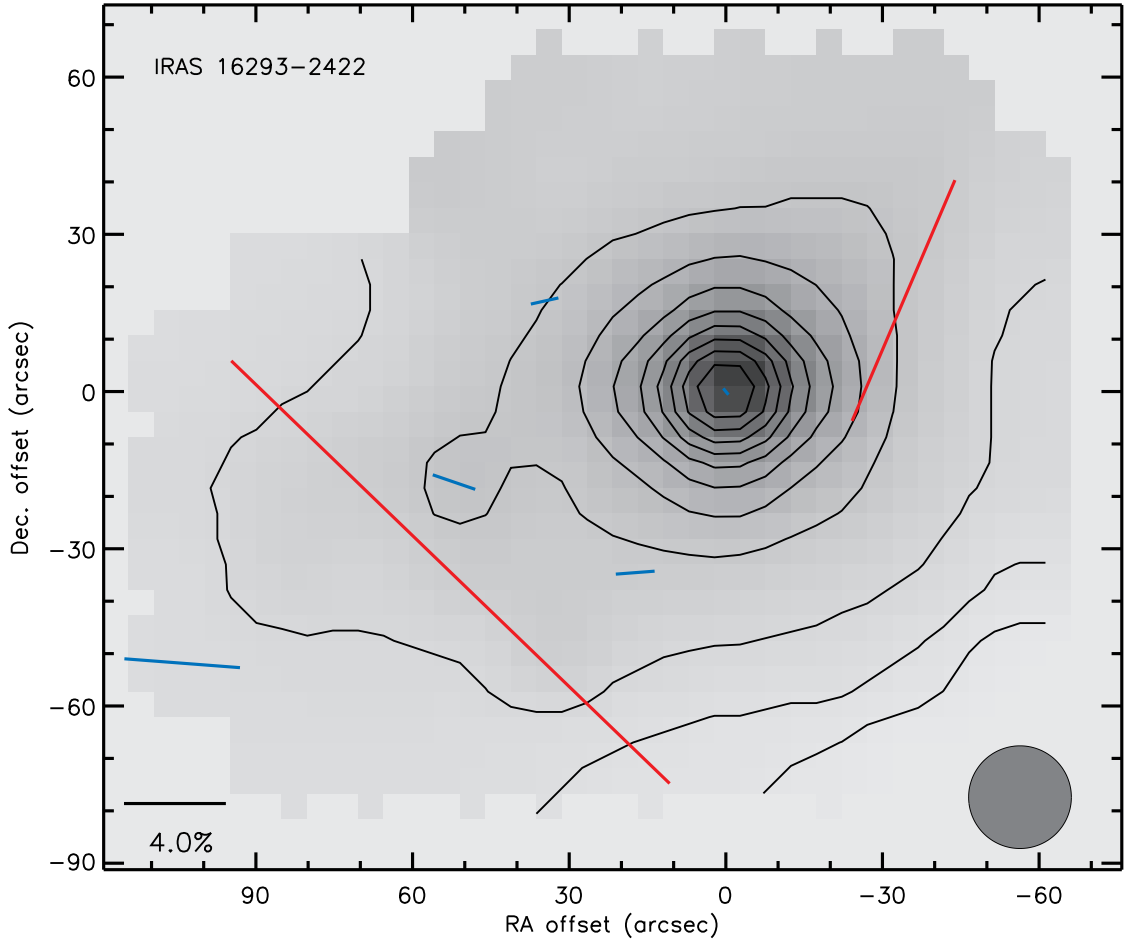}
  \caption{IRAS\,16293$-$2422.  The intensity map is from Hertz with
    contours drawn at 10, 20, 30, ..., 90\% of the peak $350\,\micron$
    intensity.}
  \label{fig-map11.5}
\end{figure}

\begin{figure}
  \includegraphics[width=0.9\columnwidth]{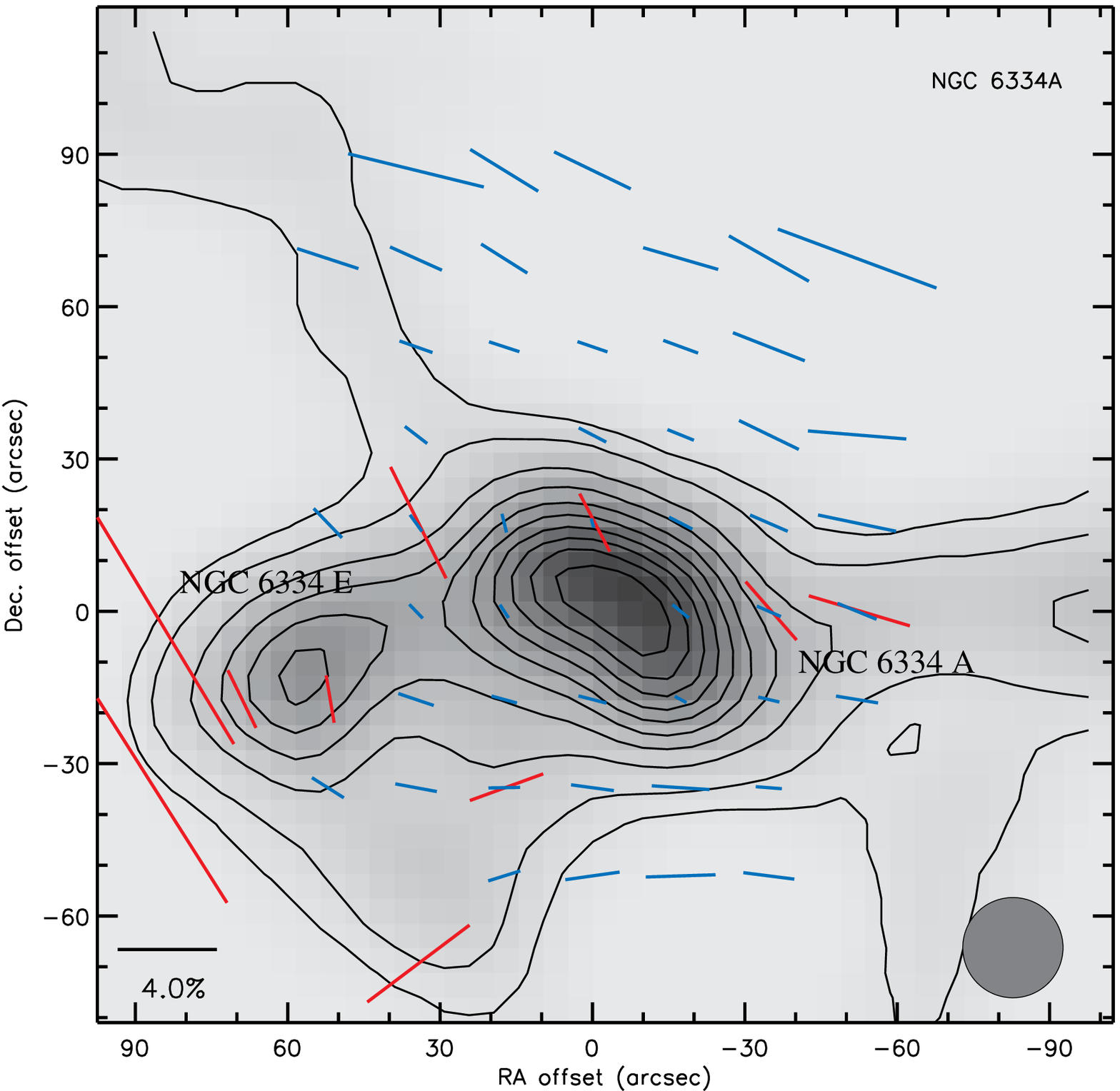}
  \caption{NGC\,6334A. The intensity map is from SCUBA with contours
    drawn at 5, 10, 20, 30, ..., 90\% of the peak $850\,\micron$
    intensity.}
  \label{fig-map12}
\end{figure}

\begin{figure}
  \includegraphics[width=0.9\columnwidth]{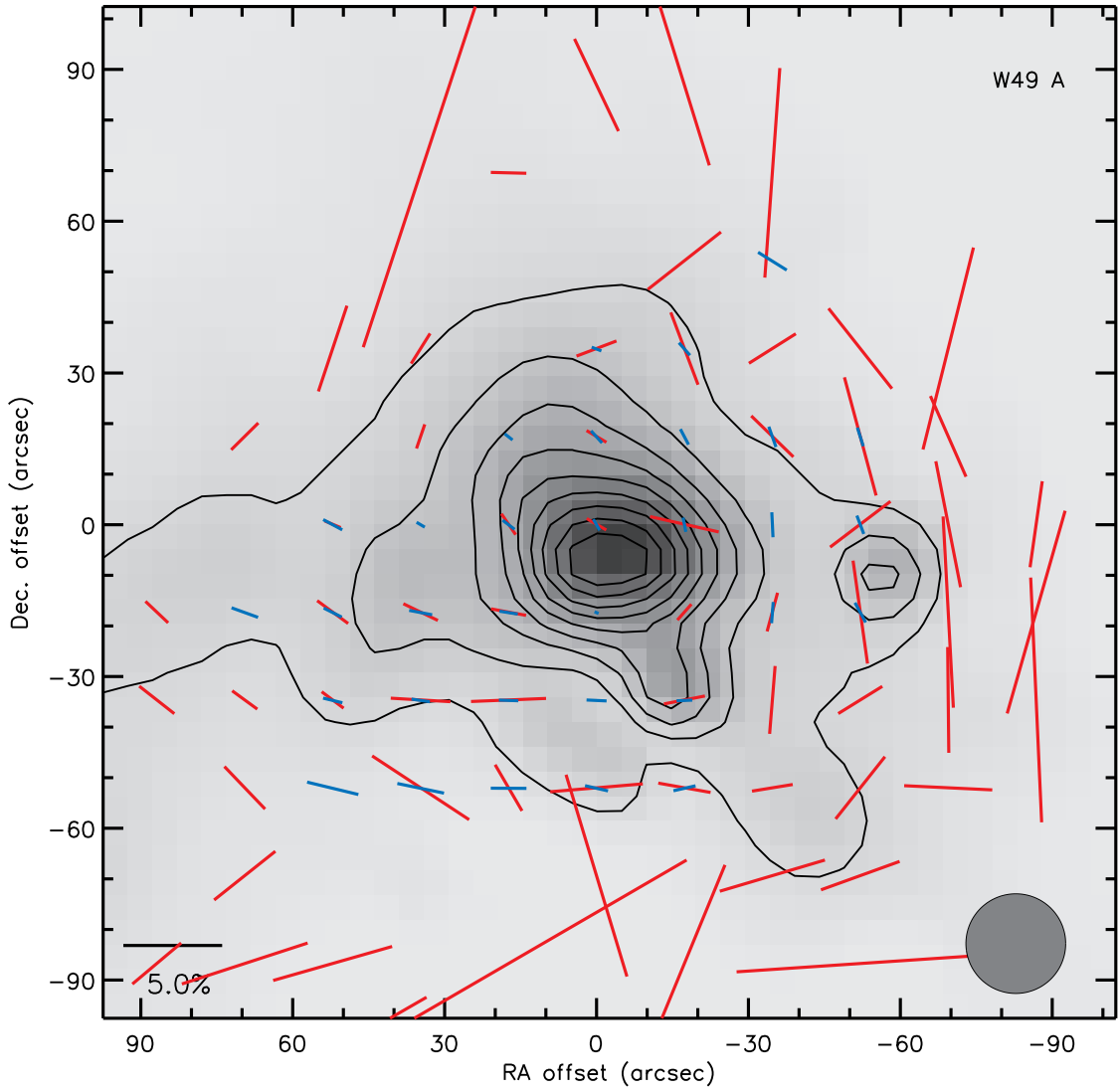}
  \caption{W49\,A. The SCUBA intensity is shown with
    contours at 10, 20, 30, ..., 90\% of the peak $850\,\micron$
    intensity.}
  \label{fig-map13}
\end{figure}

\begin{figure}
  \includegraphics[width=0.9\columnwidth]{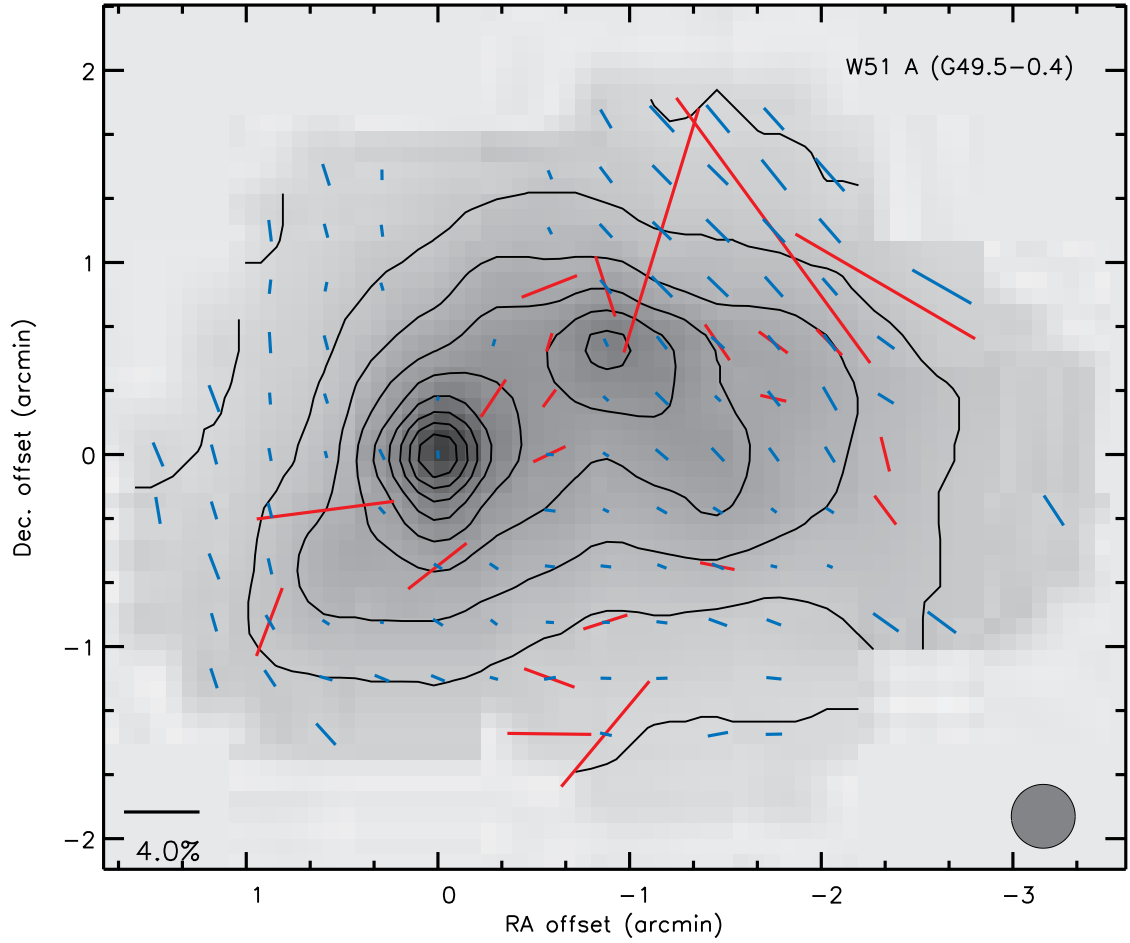}
  \caption{W51\,A (G49.5-0.4). The Hertz intensity is shown with
    contours at 10, 20, 30, ..., 90\% of the peak $350\,\micron$
    intensity.}
  \label{fig-map14}
\end{figure}

\begin{figure}
\centering
  \includegraphics[height=0.75\textheight]{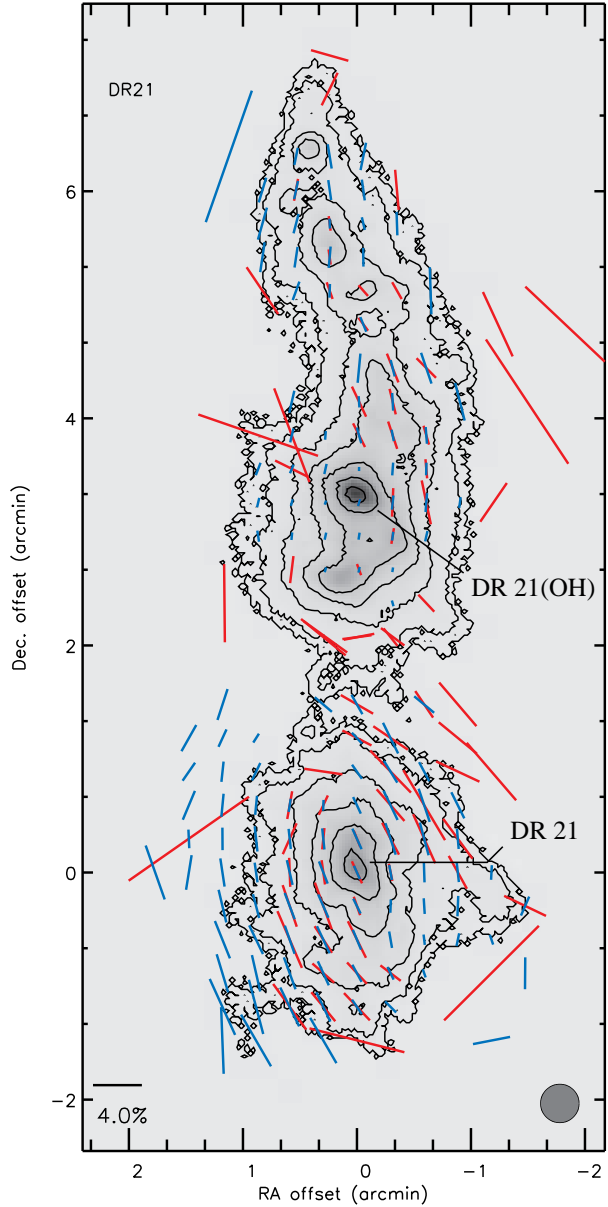}
  \caption{DR\,21.  The $350\,\micron$ intensity is shown in grayscale
    with contours drawn at 1, 2, 5, 10, 20, 40, and 80 \% of the
    peak. Intensity data are from the SHARC-2 camera with a spatial
    resolution of $\approx 10\arcsec$ (C. D. Dowell, private
    communication). The gray circle in the lower-right indicates the
    $20\arcsec$ effective beam-size of Hertz and SCUBA-pol, not the
    SHARC-2 intensity data.  The intensity peak is DR\,21 OH (Main)
    and the southern peak at the map origin is DR\,21 (Main). }
  \label{fig-map15}
\end{figure}

Figures \ref{fig-map1}--\ref{fig-map15} present grayscale/contour maps
of intensity along with polarization vectors at $350\,\micron$ (blue
lines) and $850\,\micron$ (red lines). Only vectors with $P \ge
3\sigma_p$ and $\sigma_p \le 4\%$ are drawn on the maps; the latter
criterion is an aesthetic choice to remove points with atypically
large polarizations.  Vectors spaced more closely than the nominal
$17\farcs8$ Hertz pixel pitch are a result of moving the Hertz array
in sub-pixel steps. Additionally, some SCUBA-pol vectors in regions
without Hertz coverage have not been plotted.  Each map includes a
scale-bar for determining the absolute polarization levels; the
scale-bars differ for each map but are the same length for each
wavelength within a map.  Map center coordinates are given in Table~1
of \citet{hertzarchive}.

As indicated in the figure captions most intensity maps are from SCUBA
at $850\,\micron$ \citep{scubaflux}.  For aesthetic reasons some maps
use $350\,\micron$ intensity data from Hertz, while OMC-1 and DR21 use
$350\,\micron$ intensity data from the SHARC-2 camera at the CSO
\citep{dowell2003}.  The lower-right corner of each map includes a
gray-circle indicating the $20\arcsec$ effective beam-size of Hertz
and SCUBA-pol.  Note that this beam is for the polarization data only,
not the intensity maps; SCUBA intensity maps typically have
$19\arcsec$ resolution, Hertz intensity maps typically have
$28\arcsec$ resolution, and SHARC-2 intensity maps typically have
$10\arcsec$ resolution.



\clearpage


\end{document}